\documentclass{sig-alternate}
\usepackage{mathptmx} 

\usepackage{fancyhdr}
\usepackage[normalem]{ulem}
\usepackage[hyphens,spaces,obeyspaces]{url}
\usepackage[sort,nocompress]{cite}
\usepackage[final]{microtype}
\usepackage[keeplastbox]{flushend}
\usepackage[bookmarks=true,breaklinks=true,colorlinks,linkcolor=black,citecolor=blue,urlcolor=blue]{hyperref}

\fancypagestyle{firstpage}{
  \fancyhf{}
  
  \fancyhead[C]{}
  \fancyfoot[C]{\thepage}
}

\usepackage{enumitem}
\usepackage[keeplastbox]{flushend}
\usepackage[inline]{asymptote}
\usepackage{svg}
\usepackage{subfig}
\usepackage{tabularx}
\usepackage{multirow}
\usepackage{multicol}
\usepackage{booktabs}
\usepackage{microtype}
\usepackage{siunitx}

\newcommand{\svab}[1]{}
\def\NAME{ILLIXR}

\title{Exploring Extended Reality with \NAME{}:\\ A New Playground for Architecture Research\vspace{-50pt}}
\author{Muhammad Huzaifa\quad Rishi Desai\quad Samuel Grayson\quad Xutao Jiang\quad Ying Jing\cr
Jae Lee\quad Fang Lu\quad Yihan Pang\quad Joseph Ravichandran\quad Finn Sinclair\cr
Boyuan Tian\quad Hengzhi Yuan\quad Jeffrey Zhang\quad Sarita V. Adve\cr
University of Illinois at Urbana-Champaign\cr
\texttt{illixr@cs.illinois.edu}\cr
}

\begin{document}
\maketitle
\thispagestyle{firstpage}
\pagestyle{plain}


\begin{abstract}

As we enter the era of domain-specific architectures, systems researchers must understand the requirements of emerging application domains. Augmented and virtual reality (AR/VR) or extended reality (XR) is one such important domain.
This paper presents \NAME{}, the first open source end-to-end XR system
 (1) with state-of-the-art components, (2) integrated with a modular and extensible multithreaded runtime, (3) providing an OpenXR compliant interface to XR applications (e.g., game engines), and (4) with the ability to report (and trade off) several quality of experience (QoE) metrics.
We analyze performance, power, and QoE metrics for the complete \NAME{} system and for its individual components. Our analysis reveals several properties with implications for architecture and systems research. These include demanding performance, power, and QoE requirements, a large diversity of critical tasks, 
inter-dependent execution pipelines with challenges in scheduling and resource management, and a large tradeoff space between performance/power and human perception related QoE metrics. 
\NAME{} and our analysis have the potential to propel new directions in architecture and systems research in general, and impact XR in particular.


\end{abstract}

\section{Introduction}
\label{sec:intro}

Recent years have seen the convergence of multiple disruptive trends to fundamentally change computer systems: (1) With the end of Dennard scaling and Moore's law, application-driven specialization has emerged as a key architectural technique to meet the requirements of emerging applications, (2) computing and data availability have reached an inflection point that is enabling a number of new application domains, and (3) these applications are increasingly deployed on resource-constrained edge devices, where they interface directly with the end-user and the physical world.

In response to these trends, our research conferences have seen an explosion of papers on highly efficient accelerators, many focused on machine learning. Thus, today's computer architecture researchers must not only be familiar with hardware principles, but more than ever before, must understand emerging applications. To truly achieve the promise of efficient edge computing, however, will require architects to broaden their portfolio from specialization for individual accelerators to understanding domain-specific \textit{systems} which may consist of multiple sub-domains requiring multiple accelerators that interact with each other 
to collectively meet end-user demands.


\noindent{\bf Case for Extended Reality (XR) as a driving domain:}
This paper makes the case that the emerging domain of virtual, augmented, and mixed reality, collectively referred to as extended reality (XR),\footnote{Virtual reality (VR) immerses the user in a completely digital environment. Augmented Reality (AR) enhances the user's real world with overlaid digital content. Mixed reality (MR) goes beyond AR in enabling the user to interact with virtual objects in their real world.}
is a rich domain that can propel research on efficient domain-specific edge systems. Our case rests on the following observations:\\
{\em (1) Pervasive:} XR will pervade most aspects of our lives --- it will affect the way we teach, conduct science, practice medicine, entertain ourselves, train professionals, interact socially, and more. Indeed, XR is envisioned to be the next interface for most of computing~\cite{Nvidia2018a,surgeonGame,KrohnTromp2020,KrevelenPoelman2010}. \\
{\em (2) Challenging demands:} While current XR systems exist today, they are far from providing a tetherless experience approaching perceptual abilities of humans. There is a gap of several orders of magnitude between what is needed and achievable in performance, power, and usability (Table~\ref{table:requirements}), giving architects a potentially rich space to innovate.\\ 
{\em (3) Multiple and diverse components:} XR involves a number of diverse sub-domains --- video, graphics, computer vision, machine learning, optics, audio, and components of robotics ---
making it challenging to design a system that executes each one well while respecting the resource constraints.\\
{\em (4) Full-stack implications:} The combination of real-time constraints, complex interacting pipelines, and ever-changing algorithms creates a need for full stack optimizations involving the hardware, compiler, operating system, and algorithm.\\
{\em (5) Flexible accuracy for end-to-end user experience:} The end user being a human with limited perception enables a rich space of accuracy-aware resource trade-offs, but requires the ability to quantify impact on end-to-end experience.

\noindent
{\bf Case for an XR system testbed:}
A key obstacle to architecture research for XR is that there are no open source benchmarks covering the entire XR workflow to drive such research. 
While there exist open source codes for some individual components of the XR workflow (typically developed by domain researchers), there is no integrated suite that enables researching an XR system. As we move from the era of general-purpose, homogeneous cores on chip to domain-specific, heterogeneous {\em system}-on-chip architectures, benchmarks need to follow the same trajectory. While previous benchmarks comprising of suites of independent applications (e.g., Parsec~\cite{Parsec}, Rodinia~\cite{Rodinia}, SPEC~\cite{SPEC2006a,SPEC2017}, SPLASH~\cite{SPLASH,SPLASH2,SPLASH3}, and more) sufficed to evaluate general-purpose single- and multicore architectures, there is now a need for a {\em full-system-benchmark} methodology, better viewed as a {\em full system testbed}, to design and evaluate system-on-chip architectures. Such a methodology must bring together the diversity of components that will interact with each other in the domain-specific system and also be flexible and extensible to accept future new components.

An XR full-system-benchmark or testbed will continue to enable traditional research of designing optimized accelerators for a given XR component with conventional metrics such as power, performance, and area for that component, but additionally allowing evaluations for the end-to-end impact on the system. More important, the integrated collection of components in a full XR system will enable new research that co-designs acceleration for the multiple diverse and demanding components of an XR system, across the full stack, and with end-to-end user experience as the metric. 

\noindent 
{\bf Challenges and contributions:}
This paper presents
\NAME{},
the first open-source XR full-system testbed; an analysis of performance, power, and QoE metrics for \NAME{} on a desktop class and embedded class machine; and several implications for future systems research. 

There were two challenges in the development of \NAME{}. First, \NAME{} required expertise in a large number of sub-domains (e.g., robotics, computer vision, graphics, optics, and audio). 
We developed \NAME{} after consultation with many experts in these sub-domains,
which led to identifying a representative XR workflow and
 state-of-the-art algorithms and open source codes for the constituent components.

Second, until recently, commercial XR devices had proprietary interfaces. For example, the interface between an Oculus head mounted device (HMD) runtime and the Unity or Unreal game engines that run on the HMD has been closed as are the interfaces between the different components within the HMD runtime. OpenXR~\cite{openxr}, an open standard, was released in July 2019 to partly address this problem by providing a standard for XR device runtimes to interface with the applications that run on them. \NAME{} leverages an open implementation of this new standard~\cite{monado} to  provide an OpenXR compliant XR system. However, OpenXR is still evolving, it does not address the problem of interfaces between the different components within an XR system, and the ecosystem of OpenXR compliant applications (e.g., game engines and games) is still developing. Nevertheless \NAME{} is able to provide the functionality to support sophisticated applications.

Specifically, this work makes the following contributions.

\noindent


(1) We develop \NAME{}, the first open source XR system, consisting of a representative XR workflow
 (i) with state-of-the-art components, (ii) integrated with a modular and extensible multithreaded runtime, (iii) providing an OpenXR compliant interface to XR applications (e.g., game engines), and (iv) with the ability to report (and trade off) several quality of experience (QoE) metrics.
The \NAME{} components represent the sub-domains of robotics (odometry or SLAM), computer vision (scene reconstruction), machine learning (eye tracking), image processing (camera), graphics (asynchronous reprojection), optics (lens distortion and chromatic aberration correction), audio (3D audio encoding and decoding), and displays (holographic displays).

(2) We analyze performance, power, and QoE metrics for \NAME{} on desktop and embedded class machines with CPUs and GPUs, driven by a game engine running representative VR and AR applications. Overall, we find 
that current systems are far from the needs of future devices, making the case for efficiency through techniques such as specialization, codesign, and approximation.

(3) Our system level analysis shows a wide variation in the resource utilization of different components. Although components with high resource utilization should clearly be targeted for optimization, components with relatively low  utilization are also critical due to their impact on QoE.

(4) Power breakdowns show that CPUs, GPUs, and memories are only part of the power con\-sump\-tion. The rest of the SoC and system logic, in\-clud\-ing data movement for displays and sensors is a major component as well, motivating technologies such as on-sensor computing.

(5) We find XR components are quite diverse in their use of CPU, GPU compute, and GPU graphics, and exhibit a range of IPC and system bottlenecks. Analyzing their compute and memory characteristics, we find a variety of patterns and subtasks, with none dominating. The number and diversity of these patterns poses a research question for the granularity at which accelerators should be designed and whether and how they should be shared among different components.  These observations motivate research in automated tools to identify acceleratable primitives, architectures for communication between accelerators, and accelerator software interfaces and programming models.

(6) Most components exhibit significant variability in per-frame execution time due to input-dependence or resource contention, thereby making it challenging to schedule and allocate shared resources. Further, there are a large number of system parameters that need to be tuned for an optimal XR experience. The current process is mostly ad hoc and exacerbated by the above variability.

Overall, this work provides the architecture and systems community with a one-of-a-kind infrastructure, foundational analyses, and case studies to enable new research within each layer of the system stack or co-designed across the stack, impacting a variety of domain-specific design methodologies in general and for XR in particular. \NAME{} is fully open source and is designed to be extensible so that the broader community can easily contribute new components and features and new implementations of current ones.
The current \NAME{} codebase consists of over 100K lines of code for the main components, another 80K for the OpenXR interface implementation, several XR applications and an open source game engine, and continues to grow with new components and algorithms.
\NAME{} can be found at:

\centerline{\url{https://illixr.github.io}}
\section{XR Requirements and Metrics}
\label{sec:primer}

\begin{table}[t!]
\centering
{
    \caption{%
        Ideal requirements of VR and AR vs.\ state-of-the-art devices,
        HTC Vive Pro for VR and Microsoft HoloLens 2 for AR. VR devices are typically larger and so afford more power and thermal headroom.
        The HTC Vive Pro offloads most of its work to an attached server, so its power and area values are not meaningful.  The ideal case requirements for power, area, and weight are based on current devices which are considered close to ideal in these (but not other) respects -- Snapdragon 835 in the Oculus Quest VR headset and APQ8009w in the North Focals small AR glasses.
    }
    \scriptsize
    \setlength{\tabcolsep}{3pt}
    \begin{tabular}{c c c c c}
    \textbf{Metric}                 & \textbf{HTC}          & \textbf{Ideal VR}         & \textbf{Microsoft}            & \textbf{Ideal AR} \\
                                    & \textbf{Vive Pro}     & {\bf \cite{kanter2015,ElbambyPerfecto2018}}
                                                                                        & \textbf{HoloLens 2}           & {\bf \cite{kanter2015,ElbambyPerfecto2018,takahashi2018}} \\ \toprule
    Resolution (MPixels)            & 4.6 ~\cite{vivepro}   & 200                       & 4.4~\cite{hololens2}          & 200 \\ 
    \hline
                                    & 110~\cite{vivepro}    & Full:                     & 52 diagonal                   & Full:  \\
    Field-of-view                   &                       & 165$\times$175            &  ~\cite{holowiki, wiredholo}  & 165$\times$175  \\
     (Degrees)                      &                       & Stereo:                   &                               & Stereo: \\
                                    &                       & 120$\times$135            &                               & 120$\times$135 \\ \hline
    Refresh rate (Hz)               & 90~\cite{vivepro}     & 90 -- 144                 & 120~\cite{holorefresh}        & 90 -- 144 \\ 
    \hline
    Motion-to-photon                & $<$ 20~\cite{le2018htc}
                                                            & $<$ 20                    & $<$ 9~\cite{hololatency}      & $<$ 5 \\ 
    latency (ms)                    &                       &                           &                               & \\ \hline
    Power (W)                       & N/A                   & 1 -- 2                    & $>$ 7~\cite{holoHotChips2019,Snapdragon845Power,Snapdragon855Power}
                                                                                                                        & 0.1 -- 0.2 \\ \hline
    Silicon area (mm$^2$)           & N/A                   & 100 -- 200                & $>$ 173~\cite{holoHotChips2019,Snapdragon850Area}
                                                                                                                        & $<$ 100 \\ 
    \hline
    Weight (grams)                  & 470~\cite{vivewiki}   & 100 -- 200                & 566~\cite{hololens2}          & 10s \\ 
    \end{tabular}
    \label{table:requirements}
}
\end{table}


XR requires devices that are lightweight, mobile, and all day wearable, and provide sufficient compute at low thermals and energy consumption to support a high quality of experience (QoE). 
Table~\ref{table:requirements} summarizes values for various system-level quality related metrics for current state-of-the-art XR devices and the aspiration for ideal futuristic devices. These requirements are driven by both human anatomy and usage considerations; e.g., ultra-high performance to emulate the real world, extreme thermal constraints due to contact with human skin, and light weight for comfort.
We identified the values of the various aspirational metrics through an extensive survey of the literature~\cite{kanter2015,Nvidia2018a,Nvidia2018b}. Although there is no consensus on exact figures, there is an evolving consensus on approximate values that shows orders of magnitude difference between the requirements of future devices and what is implemented today (making the exact values less important for our purpose). For example, the AR power gap alone is two orders of magnitude. Coupled with the other gaps (e.g., two orders of magnitude for resolution), the overall system gap is many orders of magnitude across all metrics.
\section{The \NAME{} System}
\label{sec:pipeline}

\begin{figure}[t]
    \centering
    \includegraphics[width=0.8\columnwidth]{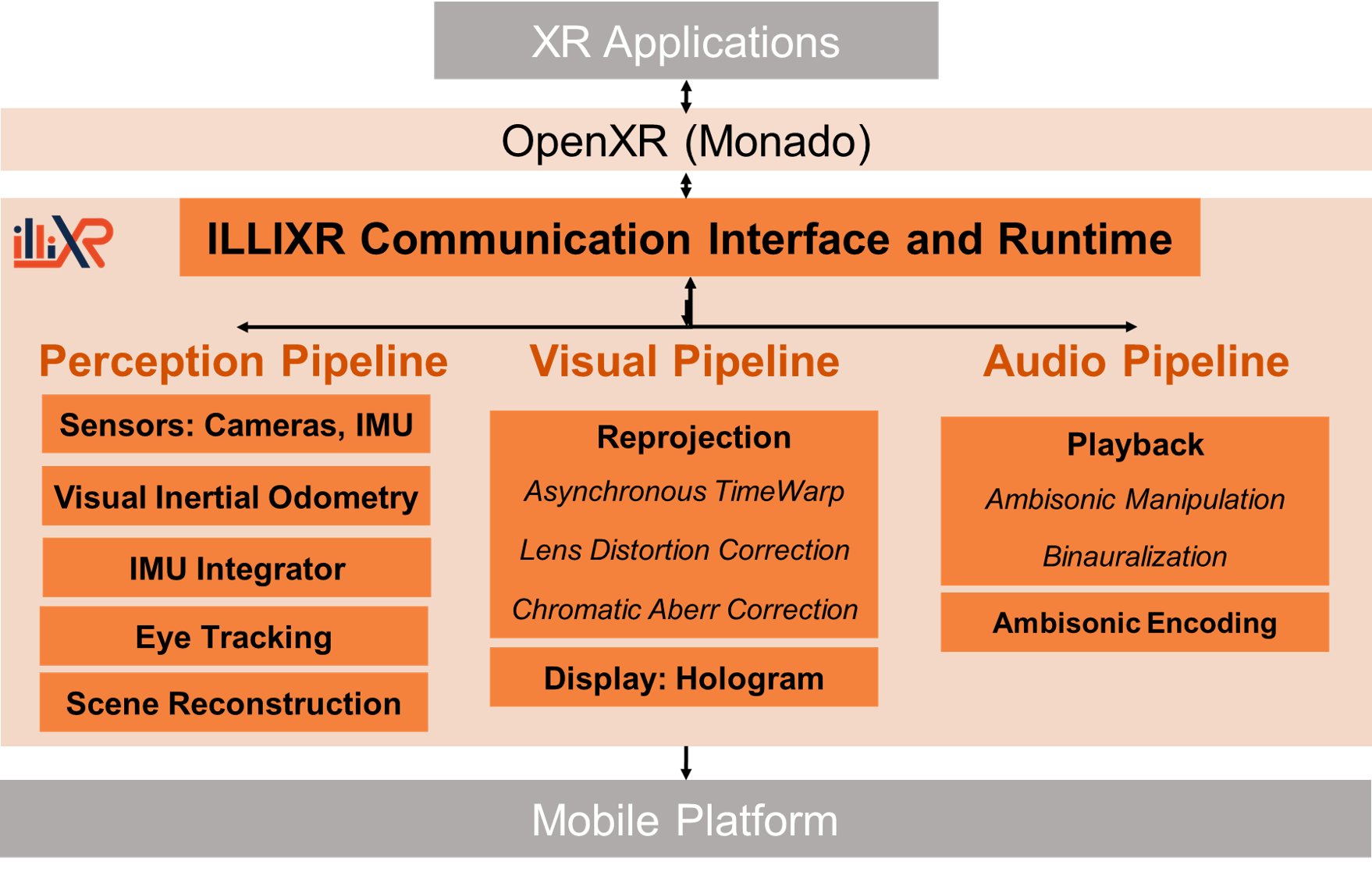}
    \caption{The \NAME{} system and its relation to XR applications, the OpenXR interface, and mobile platforms.
    }
    \label{fig:xr-system}
\end{figure}

Figure~\ref{fig:xr-system} presents the \NAME{} system. \NAME{} captures components that belong to an XR runtime such as Oculus VR (OVR) and are shipped with an XR headset. Applications, often built using a game engine, are separate from the XR system or runtime, and interface with it using the runtime's API. For example, a game developed on the Unity game engine for an Oculus headset runs on top of OVR, querying information from OVR and submitting rendered frames to it using the OVR API. Analogously, applications interface with \NAME{} using the OpenXR API~\cite{openxr} (we leverage the OpenXR API implementation from Monado~\cite{monado}). Consequently, \NAME{} does not include components from the application, but captures the performance impact of the application running on \NAME{}. We collectively refer to all application-level tasks such as user input handling, scene simulation, physics, rendering, etc. as "application" in the rest of this paper.

\begin{figure}[t!]
  \centering
  \includegraphics[width=\linewidth]{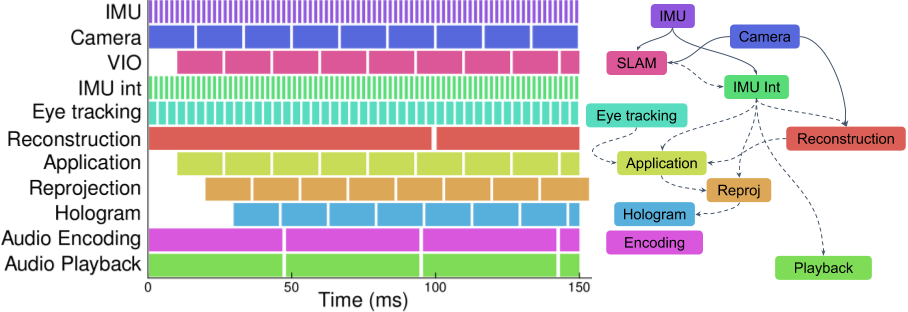}
  \caption{Interactions between \NAME{} components. The left part shows an ideal \NAME{} execution schedule and the right part shows inter-component dependencies that the \NAME{} scheduler must maintain (Section~\ref{subsec:runtime}). Solid arrows are synchronous and dashed are asynchronous dependencies.
  }
  \label{fig:comp-time}
\end{figure}

As shown in Figure~\ref{fig:xr-system}, \NAME{}'s workflow consists of three pipelines -- perception, visual, and audio -- each containing several components and interacting with each other through the \NAME{} communication interface and runtime.  Figure~\ref{fig:comp-time} illustrates these interactions -- the left side presents a timeline for an ideal schedule for the different components and the right side shows the dependencies between the different components (enforced by the \NAME{} runtime). We distilled this workflow from multiple sources that describe different parts of a typical VR, AR, or MR pipeline, including conversations with several experts from academia and industry.

\NAME{} represents a state-of-the-art system capable of running typical XR applications and providing an end-to-end XR user experience. Nevertheless, XR is an emerging and evolving domain and no XR experimental testbed or commercial device can be construed as complete in the traditional sense. Compared to specific commercial headsets today, \NAME{} may miss a component (e.g., Oculus Quest 2 provides hand tracking) and/or it may have additional or more advanced components (e.g., except for HoloLens 2, most current XR systems do not have scene reconstruction). Further, while \NAME{} supports state-of-the-art algorithms for its components, new algorithms are continuously evolving. \NAME{} therefore supports a modular and extensible design that makes it relatively easy to swap and to add new components.

The system presented here assumes all computation is done on the XR device (i.e., no offloading to the cloud, server, or other edge devices) and we assume a single user (i.e., no communication with multiple XR devices). 
Thus, our workflow does not represent any network communication. 
We are currently working on integrating support for networking, edge and cloud work partitioning, and multiparty XR within \NAME{}, but these are outside the scope of this paper.

Sections~\ref{subsec:pipeline-perception}--~\ref{subsec:pipeline-audio} next describe the three \NAME{} pipelines and their components, Section~\ref{subsec:runtime} describes the modular runtime architecture that integrates these components, and Section~\ref{subsec:metrics} describes the metrics and telemetry support in \NAME{}.
Table~\ref{table:components} summarizes the algorithm and implementation information for each component in the three pipelines. \NAME{} already supports multiple, easily interchangeable alternatives for some components; for lack of space, we pick one alternative, indicated by a * in the table, for detailed results in this paper.
Tables~\ref{table:openvins}--\ref{table:playback} summarize, for each for these components, their major tasks and sub-tasks in terms of their functionality, computation and memory patterns, and percentage contribution to execution time, discussed further below and in Section~\ref{sec:results} (we do not show the sensor related \NAME{} components and IMU integrator since they are quite simple).

\begin{table}[t!]
\centering
{
    \caption{
        \NAME{} component algorithms and implementations. GLSL stands for OpenGL Shading Language. * represents the implementation alternative for which we provide detailed results.
    }
    \scriptsize
    
    \newcolumntype{R}[1]{>{\raggedleft\let\newline\\}p{#1}}
    \newcolumntype{L}[1]{>{\raggedright\let\newline\\}p{#1}}
    \definecolor{lightgray}{gray}{0.95}
    \begin{tabularx}{\linewidth}{>{\raggedright\arraybackslash}p{2.2cm} >{\raggedright\arraybackslash}p{3.8cm} >{\raggedleft\arraybackslash}X}
    \textbf{Component}                      & \textbf{Algorithm}                    & \textbf{Implementation} \\
    \toprule
    \multicolumn{3}{l}{\textbf{Perception Pipeline}} \\
    \midrule
    Camera                                  & ZED SDK\textbf{*}~\cite{ZEDSDK}       & C++ \\
    Camera                                  & Intel RealSense SDK~\cite{IntelRealSense}
                                                                                    & C++ \\
    IMU                                     & ZED SDK~\cite{ZEDSDK}                 & C++ \\
    IMU                                     & Intel RealSense SDK~\cite{IntelRealSense}
                                                                                    & C++ \\
    VIO                                     & OpenVINS\textbf{*}~\cite{OpenVINS}    & C++ \\
    VIO                                     & Kimera-VIO~\cite{KimeraVIO}           & C++ \\
    IMU Integrator                          & RK4\textbf{*}~\cite{OpenVINS}         & C++ \\
    IMU Integrator                          & GTSAM~\cite{GTSAMRepo}                & C++ \\
    Eye Tracking                            & RITnet~\cite{ritnet}                  & Python,CUDA \\
    Scene Reconstruction                    & ElasticFusion\textbf{*}~\cite{ElasticFusion}
                                                                                    & C++,CUDA,GLSL \\
    Scene Reconstruction                    & KinectFusion~\cite{KinectFusionApp}   & C++, CUDA \\
    \midrule
    \multicolumn{3}{l}{\textbf{Visual Pipeline}} \\
    \midrule
    Reprojection                            & VP-matrix reprojection w/ pose~\cite{OculusATW}
                                                                                    & C++, GLSL \\
    Lens Distortion                         & Mesh-based radial distortion~\cite{OculusATW}
                                                                                    & C++, GLSL \\
    Chromatic Aberration                    & Mesh-based radial distortion \cite{OculusATW}
                                                                                    & C++, GLSL \\ 
    Adaptive display                        & Weighted Gerchberg–Saxton~\cite{PerssonEngstrom2011}
                                                                                    & CUDA \\ 
    \midrule
    \multicolumn{3}{l}{\textbf{Audio Pipeline}} \\
    \midrule
    Audio Encoding                          & Ambisonic encoding~\cite{libspatialaudio}
                                                                                    & C++ \\
    Audio Playback                          & Ambisonic manipulation, binauralization~\cite{libspatialaudio}
                                                                                    & C++ \\ 
    \end{tabularx}
    \label{table:components}
}
\end{table}

    \subsection{Perception Pipeline}
    \label{subsec:pipeline-perception}
    
    The perception pipeline translates the user's physical motion into information understandable to the rest of the system so it can render and play the new scene and sound for the user's new position. The input to this part comes from sensors such as cameras and an inertial measurement unit (IMU) to provide the user's acceleration and angular velocity. The processing components include camera and IMU processing, head tracking or VIO (Visual Inertial Odometry) for obtaining low frequency but precise estimates of the user's pose (the position and orientation of their head), IMU integration for obtaining high frequency pose estimates, eye tracking for determining the gaze of the user, and scene reconstruction for generating a 3D model of the user's surroundings. We elaborate on some of the key components below.
        
        \subsubsection{Camera}
        \label{subsubsec:pipeline-perception-camera}
        
        \NAME{} can work with any commercial camera. We provide support for both ZED Mini and Intel RealSense cameras. For this paper, we use a ZED Mini~\cite{ZEDMini} as it provides IMU readings, stereo RGB images, and depth images all together with a simple API.
        The camera component first asks the camera to obtain new stereo camera images and then copies these images into main memory.
        
        \subsubsection{IMU}
        \label{subsubsec:pipeline-perception-imu}
        
        \NAME{}'s IMU component queries the IMU at fixed intervals to obtain linear acceleration from the accelerometer and angular velocity from the gyroscope. In addition to retrieving the IMU values, this component converts the angular velocity from $deg/sec$ to $rad/sec$, and timestamps the readings for use by other components, such as VIO and IMU integrator.
        
        \subsubsection{VIO}
        \label{subsubsec:pipeline-perception-vio}
        
        Head tracking is achieved through Simultaneous Localization and Mapping (SLAM), a technique that can track the user's motion without the use of external markers in the environment, and provide both the position and orientation of the user's head; i.e., the six degrees of freedom or 6DOF pose of the user.
        There are a large number of algorithms that have been proposed for SLAM for a variety of scenarios, including for robots, drones, VR, and AR. We support both OpenVINS~\cite{openVINS2019iros} and Kimera-VIO~\cite{Rosinol20icra-Kimera}. In this paper, we use OpenVINS as it has been shown to be more accurate than other popular algorithms such as VINS-Mono, VINS-Fusion, and OKVIS~\cite{openVINS2019iros}. Both OpenVINS and Kimera-VIO do not have mapping capabilities, and are therefore called Visual-Inertial Odometry (VIO) algorithms.
        Table~\ref{table:openvins} overviews OpenVINS by describing its tasks.\footnote{Due to space constraints, we are unable to describe here the details of the algorithms in the components of \NAME{}.
        }
        
        \subsubsection{IMU Integrator}
        \label{subsubsec:pipeline-perception-integrator}
        
        VIO only generates poses at the rate of the camera, which is typically low. To obtain high speed estimates of the user's pose, an IMU integrator integrates incoming IMU samples from the IMU to obtain a relative pose and adds it to the VIO's latest pose to obtain the current pose of the user.
        \NAME{} has two IMU integrators available: one utilizes GTSAM to perform integration~\cite{GTSAMFactorGraphs,GTSAMPreintegration} and the other uses RK4, inspired by OpenVINS' IMU integrator\cite{openVINS2019iros}. We use the RK4 integrator in this paper.

        \subsubsection{Eye Tracking}
        \label{subsubsec:pipeline-perception-eyetracking}
        
        We used RITnet for eye-tracking~\cite{ritnet} as it won the OpenEDS eye tracking challenge~\cite{openeds} with an accuracy of 95.3\%.
        RITnet is a deep neural network that combines U-Net~\cite{unet} and DenseNet~\cite{huang2016densely}.
        The network takes as input a grayscale eye image and segments it into the background, iris, sclera, and pupil. Since neural networks are well understood in the architecture community, we omit the task table for space.
        
        \subsubsection{Scene Reconstruction}
        \label{subsubsec:pipeline-perception-reconstruction}
        
        For scene reconstruction, we chose KinectFusion~\cite{NewcombeIzadi2011} and ElasticFusion~\cite{whelan2015elasticfusion}, two dense SLAM algorithms that use an RGB-Depth (RGB-D) camera to build a dense 3D map of the world.
        In this paper, we focus on ElasticFusion (Table~\ref{table:elasticfusion}), which uses a surfel-based map representation, where each map point, called a surfel, consists of a position, normal, color, radius, and timestamp.
        In each frame, incoming color and depth data from the camera is used to update the relevant surfels in the map.
        
    \begin{table}[t!]
\centering
{
    \caption{
        Task breakdown of VIO.
    }
    \scriptsize
    \newcolumntype{L}[1]{>{\raggedright\let\newline\\}p{#1}}
    \renewcommand{\tabularxcolumn}[1]{>{\raggedright\arraybackslash}p{#1}}
    
    \begin{tabularx}{\linewidth}{L{2.0cm} L{0.3cm} L{2.5cm} X}
    
    \textbf{Task}       & \textbf{Time}     & \textbf{Computation}      & \textbf{Memory Pattern} \\
    \toprule
    
    \textbf{Feature detection} \newline Detects new features in the new camera images
                        & 15\%              & Integer stencils per pyramid level
                                                                        & Locally dense stencil; globally mixed dense and sparse \\
    \midrule
    
    \textbf{Feature matching} \newline Matches features across images
                        & 13\%              & Integer stencils; GEMM; linear algebra
                                                                        & Locally dense stencil; globally mixed dense and sparse; mixed dense and random feature map accesses \\
    \midrule
    
    \textbf{Feature initialization} \newline Adds new features to state
                        & 14\%              & SVD; Gauss-Newton; Jacobian; nullspace projection; GEMM
                                                                        & Dense feature map accesses; mixed dense and sparse state matrix accesses \\
    \midrule
    
    \textbf{MSCKF update} \newline Updates state using MSCKF algorithm
                        & 23\%              & SVD; Gauss-Newton; Cholesky; QR; Jacobian; nullspace projection; chi$^2$ check; GEMM
                                                                        & Dense feature map accesses; mixed dense and sparse state matrix accesses \\
    \midrule
    
    \textbf{SLAM update} \newline Updates state using EKF-SLAM algorithm
                        & 20\%              & Identical to \textbf{MSCKF update}
                                                                        & Similar, but not identical, to \textbf{MSCKF update} \\
    \midrule
    
    \textbf{Marginalization} \newline Removes features from state
                        & 5\%               & Cholesky; matrix arithmetic
                                                                        & Dense feature map and state matrix accesses \\
    \midrule
    
    \textbf{Other} \newline  Miscellaneous tasks
                        & 10\%              & Gaussian filter; histogram
                                                                        & Globally dense stencil \\
    \end{tabularx}
    \label{table:openvins}
}
\end{table}

    \begin{table}[t!]
\centering
{
    \caption{
        Task breakdown of scene reconstruction.
    }
    \scriptsize
    \newcolumntype{L}[1]{>{\raggedright\let\newline\\}p{#1}}
    \renewcommand{\tabularxcolumn}[1]{>{\raggedright\arraybackslash}p{#1}}
    
    \begin{tabularx}{\linewidth}{L{2.3cm} L{0.3cm} L{2.7cm} X }
    
    \textbf{Task}       & \textbf{Time}    & \textbf{Computation}      & \textbf{Memory Pattern} \\
    \midrule
    
    \textbf{Camera Processing} \newline Processes incoming camera depth image
                        & 5\%               & Bilateral filter; invalid depth rejection
                                                                                                & Dense sequential accesses to depth image \\
    \midrule
    
    \textbf{Image Processing} \newline Pre-processes RGB-D image for tracking and mapping
                        & 18\%              & Generation of vertex map, normal map, and image intensity; image undistortion; pose transformation of old map
                                                                                                & Globally dense; local stencil; layout change from RGB\_RGB $\rightarrow$ RR\_GG\_BB; \\

    \midrule
    
    \textbf{Pose Estimation} \newline Estimates 6DOF pose
                        & 28\%              & ICP; photometric error; geometric error
                                                                                                & photometric error is globally dense; others are globally sparse, locally dense \\
    \midrule
    
    \textbf{Surfel Prediction} \newline Calculates active surfels in current frame
                        & 38\%              & Vertex and fragment shaders
                                                                                            & Globally sparse; locally dense \\
    \midrule
    
    \textbf{Map Fusion} \newline Updates map with new surfel information
                        & 11\%              & Vertex and fragment shaders
                                                                                            & Globally sparse; locally dense \\
    
    \end{tabularx}
    \label{table:elasticfusion}
}
\end{table}

    \subsection{Visual Pipeline}
    \label{subsec:pipeline-visual}
    
    \begin{table}[t!]
\centering
{
    \caption{
        Task breakdown of reprojection.
    }
    \scriptsize
    \newcolumntype{L}[1]{>{\raggedright\let\newline\\}p{#1}}
    \renewcommand{\tabularxcolumn}[1]{>{\raggedright\arraybackslash}p{#1}}
    \begin{tabularx}{\linewidth}{p{2.3cm} c X X }
    
    \textbf{Task}       & \textbf{Time}     & \textbf{Computation}      & \textbf{Memory Pattern} \\
    \midrule
    
    \textbf{FBO} \newline FBO state management
                        & 24\%              & Framebuffer bind and clear
                                                                        & Driver calls; CPU-GPU communication \\
    \midrule
    
    \textbf{OpenGL State Update} \newline Sets up OpenGL state
                        & 54\%              & OpenGL state updates; one drawcall per eye
                                                                        & Driver calls; CPU-GPU communication \\
    \midrule
    
    \textbf{Reprojection} \newline Applies reprojection transformation to image
                        & 22\%              & 6 matrix-vector MULs/vertex
                                                                        & Accesses uniform, vertex, and fragment buffers; 3 texture accesses/fragment \\
    \end{tabularx}
    \label{table:timewarp}
}
\end{table}

    \begin{table}[t!]
\centering
{
    \caption{
        Task breakdown of hologram.
    }
    \scriptsize
    \newcolumntype{L}[1]{>{\raggedright\let\newline\\}p{#1}}
    \renewcommand{\tabularxcolumn}[1]{>{\raggedright\arraybackslash}p{#1}}
    \begin{tabularx}{\linewidth}{L{2.1cm} p{.7cm} L{1.75cm} X }

    \textbf{Task}       & \textbf{Time}     & \textbf{Computation}      & \textbf{Memory Pattern} \\
    \midrule
    
    \textbf{Hologram-to-depth} Propagates pixel phase to depth plane
                        & 57\%              & Transcendentals; FMADDs; TB-wide tree reduction
                                                                        & Dense row-major; spatial locality in pixel data; temporal locality in depth data; reduction in scratchpad \\
    \midrule
    
    \textbf{Sum} Sums phase differences from hologram-to-depth
                        & $<$ 0.1\%         & Tree reduction            & Dense row-major; reduction in scratchpad \\
    \midrule
    
    \textbf{Depth-to-hologram} Propagates depth plane phase to pixel
                        & 43\%              & Transcendentals; FMADDs; thread-local reduction
                                                                        & Dense row-major; no pixel reads; pixels written once \\
    
    \end{tabularx}
    \label{table:hologram}
}
\end{table}
    
    The visual pipeline takes information about the user's new pose from the perception pipeline, the submitted frame from the application, and produces the final display using two parts: asynchronous reprojection and display.
    
        \subsubsection{Asynchronous Reprojection}
        \label{subsubsec:pipeline-visual-reprojection}
        
        Asynchronous reprojection corrects the rendered image submitted by the application for optical distortions and compensates for latency from the rendering process (which adds a lag between the user's movements and the delivery of the image to the user).
        The reprojection component reprojects the latest available frame based on a prediction of the user's movement, resulting in less perceived latency. The prediction is performed by obtaining the latest pose from the IMU integrator and extrapolating it to the display time of the image using a constant acceleration model.
        Reprojection is also used to compensate for frames that might otherwise be dropped because the application could not submit a frame on time to the runtime. Reprojection is critical in XR applications, as any perceived latency will cause discomfort~\cite{daqriATW,nvidiaWarp,OG}.
        
        In addition to latency compensation, the reprojection component performs lens distortion correction and chromatic aberration correction. Most XR devices use small displays placed close to the user's eyes; there\-fore, some optical solution is required to create a comfortable focal distance for the user~\cite{koreDistortion}. These optics usually induce significant optical distortion and chromatic aberration -- lines become curved and colors are non-uniformly refracted. These optical affects are corrected by applying reverse distortions to the rendered image, such that the distortion from the optics is mitigated~\cite{OculusATW}.
        
        Our reprojection, lens distortion correction, and chromatic aberration correction implementation is derived from~\cite{OculusATWgit}, which only supports rotational reprojection.\footnote{We are currently developing positional reprojection.}
        We extracted the distortion and aberration correction, and reprojection shaders from this reference, and implemented our own version in OpenGL that integrates reprojection, lens distortion correction, and chromatic aberration correction into one component (Table~\ref{table:timewarp}).
        
        \subsubsection{Display}
        \label{subsubsec:pipeline-visual-display}
        
        The fixed focal length of modern display optics causes a vergence-accommodation conflict (VAC), where the user's eyes converge at one depth, which varies with the scene, and the eye lenses focus on another, which is fixed due to the display optics. VAC is a common cause of headaches and fatigue in modern XR devices~\cite{Kramida2016,ChangKumar2018,PadmanabanKonrad2017,TurnbullPhillips2017}. One possible solution to this problem is computational holography, wherein a phase change is applied to each pixel using a Spatial Light Modulator~\cite{maimone2017} to generate several focal points instead of just one. Since humans cannot perceive depths less than 0.6 diopters apart~\cite{LiuHua2010}, typically 10 or so depth planes are sufficient. The per-pixel phase mask is called a hologram.
        
        In \NAME{}, we use Weighted Gerchberg–Saxton (GSW)~\cite{LeonardoIanni2007} to generate holograms (Table~\ref{table:hologram}). We used a reference CUDA implementation of GSW~\cite{PerssonEngstrom2011}, and extended it to support RGB holograms of arbitrary size.

    \subsection{Audio Pipeline}
    \label{subsec:pipeline-audio}
    
    \begin{table}[t!]
\centering
{
    \caption{
        Task breakdown of audio encoding.
    }
    \scriptsize
    \newcolumntype{L}[1]{>{\raggedright\let\newline\\}p{#1}}
    \renewcommand{\tabularxcolumn}[1]{>{\raggedright\arraybackslash}p{#1}}
    
    \begin{tabularx}{\linewidth}{L{3.1cm} c X r }
    
    \textbf{Task}       & \textbf{Time}      & \textbf{Computation}      & \textbf{Memory Pattern} \\
    \midrule
    
    \textbf{Normalization} \newline INT16 to FP32
                        & 7\%               & Element-wise FP32 division
                                                                                                                    & Dense row-major \\
    \midrule
    \textbf{Encoding} \newline Sample to soundfield mapping
                        & 81\%              & $Y[j][i] = D \times X[j]$
                                                                                                                    & Dense column-major \\
    \midrule
    
    \textbf{Summation} \newline HOA soundfield summation   & 11\%              & $Y[i][j]+=X_k[i][j]$ $\forall k$
                                                                                                                    & Dense row-major \\
    \end{tabularx}
    \label{table:recording}
}
\end{table}
    \begin{table}[t!]
\centering
{
    \caption{
        Task breakdown of audio playback.
    }
    \scriptsize
    \newcolumntype{L}[1]{>{\raggedright\let\newline\\}p{#1}}
    \renewcommand{\tabularxcolumn}[1]{>{\raggedright\arraybackslash}p{#1}}
    
    \begin{tabularx}{\linewidth}{L{2.0cm} L{0.3cm} L{2.4cm} X}
    
    \textbf{Task}       & \textbf{Time}     & \textbf{Computation}  & \textbf{Memory Pattern} \\
    \toprule
    
    \textbf{Psychoacoustic filter} \newline Applies optimization filter
                        & 29\%              & FFT; frequency domain convolution; IFFT
                                                                    & Butterfly pattern; dense row-major sequential accesses \\
    \midrule
    
    \textbf{Rotation} \newline Rotates soundfield using pose
                        & 6\%               & Transcendentals; FMADDs
                                                                    & Sparse column-major accesses; some temporal locality \\
    \midrule
    
    \textbf{Zoom} \newline Zooms soundfield using pose
                        & 5\%               & FMADDs                & Dense column-major sequential accesses \\
    \midrule
    
    \textbf{Binauralization} \newline Applies HRTFs
                        & 60\%              & Identical to \textbf{psychoacoustic filter}
                                                                    & Identical to \textbf{psychoacoustic filter} \\
    \end{tabularx}
    \label{table:playback}
}
\end{table}
    
    The audio pipeline is responsible for generating realistic spatial audio and is composed of audio encoding and playback (Table~\ref{table:recording} and Table~\ref{table:playback}). Our audio pipeline uses libspatialaudio, an open-source Higher Order Ambisonics library~\cite{libspatialaudio}.
    
        \subsubsection{Encoding}
        \label{subsubsec:pipeline-audio-encoding}
        
        We perform spatial audio encoding by encoding several different monophonic sound sources simultaneously into an Ambisonic soundfield~\cite{Hollerweger2008}.
        
        \subsubsection{Playback}
        \label{subsubsec:pipeline-audio-playback}
        
        During playback, Ambisonic soundfield manipulation uses the user's pose (obtained from the IMU integrator) to rotate and zoom the soundfield. Then, binauralization maps the manipulated soundfield to the available speakers (we assume headphones, as is common in XR). Specifically, it applies Head-Related Transfer Functions (HRTFs)~\cite{frank2015}, which are digital filters to capture the modification of incoming sound by the person's nose, head, and outer ear shape. Adding such cues allows the digitally rendered sound to mimic real sound, helping the brain localize sounds akin to real life. There are separate HRTFs for the left and right ears due to the asymmetrical shape of the human head.

\subsection{Runtime}
\label{subsec:runtime}

Figure~\ref{fig:comp-time} shows an ideal execution timeline for the different XR components above and their temporal dependencies. On the left, each colored rectangle boundary represents the period of the respective component --- ideally the component would finish execution before the time for its next invocation. The right side shows a static representation of the dependencies among the components, illustrating the interaction between the different pipelines.

We say a consumer component exhibits a {\em synchronous} dependence on a producer component if the former has to wait for the last invocation of the latter to complete. An {\em asynchronous} dependence is softer, where the consumer can start execution with data from a previously complete invocation of the producer component. For example, the application is asynchronously dependent on the IMU integrator and VIO is synchronously dependent on the camera. The arrows in Figure~\ref{fig:comp-time} illustrate these inter-component dependencies, with solid indicating synchronous dependencies and dashed asynchronous.

An XR system is unlikely to follow the idealized schedule in Figure~\ref{fig:comp-time} (left) due to shared and constrained resources and variable running times. Thus, an explicit runtime is needed for effective resource management and scheduling while maintaining the inter-component dependencies, resource constraints, and quality of experience.

The \NAME{} runtime schedules resources while enforcing dependencies, in part deferring to the underlying Linux scheduler and GPU driver. Currently, asynchronous components are scheduled by launching persistent threads that wake up at desired intervals (thread wakes are implemented via \texttt{sleep} calls). Synchronous components are scheduled whenever a new input arrives. We do not preempt invocations of synchronous components that miss their target deadline (e.g., VIO will process all camera frames that arrive). However, asynchronous components can miss work if they do not meet their target deadline (e.g., if timewarp takes too long, the display will not be updated).
To achieve extensibility, \NAME{} is divided into a communication framework, a set of plugins, and a plugin-loader.

The communication framework is structured around {\em event streams}. Event streams support writes, asynchronous reads, and synchronous reads. Synchronous reads allow a consumer to see every value produced by the producer, while asynchronous reads allow a consumer to ask for the latest value.

Plugins are shared-object files to comply with a small interface. They can only interact with other plugins through the provided communication framework. This architecture allows for modularity. Each of the components described above is incorporated into its own plugin.

The plugin loader can load a list of plugins defined at runtime. A plugin is completely interchangeable with another as long as it writes to the same event streams in the same manner. Future researchers can test alternative implementations of a single plugin without needing to reinvent the rest of the system. Development can iterate quickly, because plugins are compiled independently and linked dynamically.

\NAME{} can plug into the Monado OpenXR runtime~\cite{monado} to support OpenXR applications.
While not every OpenXR feature is implemented, we support pose queries and frame submissions, which are sufficient for the rest of the components to function.
This allows \NAME{} to operate with many OpenXR-based applications, including those developed using game engines such as Unity, Unreal, and Godot (currently only Godot has OpenXR support on Linux, our supported platform).
However, OpenXR (and Monado) is not required to use \NAME{}. Applications can also interface with \NAME{} directly, without going through Monado, enabling \NAME{} to run standalone.


\subsection{Metrics}
\label{subsec:metrics}

The \NAME{} framework provides a multitude of metrics to evaluate the quality of the system. In addition to reporting standard performance metrics such as per-component frame rate and execution time, \NAME{} reports several quality metrics: 1) motion-to-photon latency~\cite{daqriATW}, a standard measure of the lag between user motion and image updates; 2) Structural Similarity Index Measure (SSIM)~\cite{wang2004image}, one of the most commonly used image quality metrics in XR studies~\cite{MengPaul2020,LaiHu2019,SchollmeyerSchneegans2017}; 3) FLIP~\cite{AnderssonPontus2020}, a recently introduced image quality metric that addresses the shortcomings of SSIM; and 4) pose error~\cite{ZhangScaramuzza2018}, a measure of the accuracy of the poses used to render the displayed images. The data collection infrastructure of \NAME{} is generic and extensible, and allows the usage of any arbitrary image quality metric. We do not yet compute a quality metric for audio beyond bitrate, but plan to add the recently developed AMBIQUAL~\cite{NarbuttSkoglund2020} in the future.




Overall, we believe the runtime described in this paper gives researchers a jumping-point to start answering questions related to scheduling and other design decisions that trade off various QoE metrics with conventional system constraints such as PPA.
\section{Experimental Methodology}
\label{sec:implementation}

The goals of our experiments are to quantitatively show that (1) XR presents a rich opportunity for computer architecture and systems research in domain-specific edge systems; (2) fully exploiting this opportunity requires an end-to-end system that models the complex, interacting pipelines in an XR workflow, and (3) \NAME{} is a unique testbed that provides such an end-to-end system, enabling new research directions in domain-specific edge systems. 

Towards the above goals, we perform a comprehensive characterization of a live end-to-end \NAME{} system on multiple hardware platforms with representative XR workloads. We also characterize standalone components of \NAME{} using appropriate component-specific off-the-shelf datasets. This section details our experimental setup.

\subsection{Experimental Setup}
\label{subsec:implementation-platform}

There are no existing systems that meet all the aspirational performance, power, and quality criteria for a fully mobile XR experience as summarized in Table~\ref{table:requirements}. For our characterization, we chose to run \NAME{} on the following two hardware platforms, with three total configurations, representing a broad spectrum of power-performance tradeoffs and current XR devices (the CUDA and GLSL parts of our components as shown in Table~\ref{table:components} run on the GPU and the rest run on the CPU).

\noindent
\textbf{A high-end desktop platform:} We used a state-of-the-art desktop system with an Intel Xeon E-2236 CPU (6C12T) and a discrete NVIDIA RTX 2080 GPU. This platform's thermal design power (TDP) rating is far above what is deemed acceptable for a mobile XR system, but it is representative of the platforms on which current tethered systems run (e.g., HTC Vive Pro as in Table~\ref{table:requirements}) and can be viewed as an approximate upper bound for performance of CPU+GPU based near-future embedded (mobile) systems.

\noindent
\textbf{An embedded platform with two configurations:} We used an NVIDIA Jetson AGX Xavier development board~\cite{Xavier} consisting of an Arm CPU (8C8T) and an NVIDIA Volta GPU. All experiments were run with the Jetson in 10 Watt mode, the lowest possible preset. We used two different configurations -- a high performance one (\emph{Jetson-HP}) and a low power one (\emph{Jetson-LP}). We used the maximum available clock frequencies for Jetson-HP and half those for Jetson-LP.
These configurations approximate the hardware and/or the TDP rating of several commercial mobile XR devices. For example, Magic Leap One~\cite{MagicLeapOne} used a Jetson TX2~\cite{TX2}, which is similar in design and TDP to our Xavier configuration. HoloLens 2~\cite{holoHotChips2019,Snapdragon845Power,Snapdragon855Power} and the Qualcomm Snapdragon 835~\cite{Snapdragon835Power} used in Oculus Quest~\cite{OculusQuest} have TDPs in the same range as our two Jetson configurations.

Our I/O setup is as follows. For the perception pipeline, we connected a ZED Mini camera~\cite{ZEDMini} to the above platforms via a USB-C cable. A user walked in our lab with the camera, providing live camera and IMU input (Table~\ref{table:components}) to \NAME{}. 
For the visual pipeline, we run representative VR and AR applications on a game engine on \NAME{} (Section~\ref{subsec:implementation-applications}) -- these applications interact with the perception pipeline to provide the visual pipeline with the image frames to display.

We currently display the (corrected and reprojected) images on a desktop LCD monitor connected to the above hardware platforms. We use a desktop monitor due its ability to provide multiple resolution levels and refresh rates for experimentation. Although this means that the user does not see the display while walking with the camera, we practiced a trajectory that provides a reasonable response to the displayed images and used that (live) trajectory to collect results.\footnote{We plan to support North Star~\cite{NorthStar} to provide an HMD to the user.}
Finally, for the audio pipeline, we use pre-recorded input (see Section~\ref{subsec:implementation-integration}).

We run our experiments for approximately 30 seconds. There is some variability due to the fact that these are real user experiments, and the user may perform slightly different actions based on application content.
    
\subsection{Integrated \NAME{} System Configuration}
\label{subsec:implementation-integration}

The end-to-end integrated \NAME{} configuration in our experiments uses the components as described in Table~\ref{table:components} except for scene reconstruction, eye tracking, and hologram. The OpenXR standard currently does not support an interface for an application to use the results of scene reconstruction and only recently added an interface for eye tracking. We, therefore, do not have any applications available to use these components in an integrated setting. Although we can generate holograms, we do not yet have a holographic display setup with SLMs and other optical elements, and there are no general purpose pre-assembled off-the-shelf holographic displays available. We do report results in standalone mode for these components using off-the-shelf component-specific datasets.

\begin{table}[t!]
\centering
{
    \caption{
        Key \NAME{} parameters that required manual system-level tuning. Several other parameters were tuned at the component level.
    }
    \scriptsize
    
    \begin{tabular}{l l l l}
    \textbf{Component}          & \textbf{Parameter Range}          & \textbf{Tuned}    & \textbf{Deadline} \\ \toprule
    Camera (VIO)                & Frame rate = 15 -- 100 Hz         & 15 Hz             & 66.7 ms \\
                                & Resolution = VGA -- 2K            & VGA               & -- \\
                                & Exposure = 0.2 -- 20 ms           & 1 ms              & -- \\ \hline
    IMU (Integrator)            & Frame rate = $\leq$800 Hz         & 500 Hz            & 2 ms \\ \hline
    Display (Visual pipeline,   & Frame rate = 30 -- 144 Hz         & 120 Hz            & 8.33 ms \\
    Application)                & Resolution = $\leq$2K             & 2K                & -- \\
                                & Field-of-view = $\leq$ \ang{180}  & \ang{90}          & -- \\ \hline
    Audio (Encoding,            & Frame rate = 48 -- 96 Hz          & 48 Hz             & 20.8 ms \\
    Playback)                   & Block size = 256 -- 2048          & 1024              & --
    \end{tabular}
    \label{table:parameters}
}
\end{table}

Configuring an XR system requires tuning multiple parameters of the different components to provide the best end-to-end user experience on the deployed hardware. This is a complex process involving the simultaneous optimization of many QoE metrics and system parameters. Currently tuning such parameters is a manual, mostly ad hoc process. Table~\ref{table:parameters} summarizes the key parameters for \NAME{}, the range available in our system for these parameters, and the final value we chose at the end of our manual tuning. We made initial guesses based on our intuition, and then chose final values based on both our perception of the smoothness of the system and profiling results. We expect the availability of \NAME{} will enable new research in more systematic techniques for performing such end-to-end system optimization.

\subsection{Applications}
\label{subsec:implementation-applications}

To evaluate \NAME{}, we used four different XR applications: Sponza~\cite{sponza}, Materials~\cite{material}, Platformer~\cite{platformer}, and a custom AR demo application with sparse graphics. The AR application contains a single light source, a few virtual objects, and an animated ball. We had to develop our own AR application because existing applications in the XR ecosystem all predominantly target VR (outside of Microsoft HoloLens). The applications were chosen for diversity of rendering complexity, with Sponza being the most graphics-intensive and AR demo being the least. We did not evaluate an MR application as OpenXR does not yet contain a scene reconstruction interface.

\NAME{} is currently only available on Linux, and since Unity and Unreal do not yet have OpenXR support on Linux, all four applications use the Godot game engine~\cite{godot}, a popular open-source alternative with Linux OpenXR support.

\subsection{Experiments with Standalone Components}
\label{subsec:implementation-datasets}

For more detailed insight into the individual components of \NAME{}, we also ran each component by itself on our desktop and embedded platforms using off-the-shelf standalone datasets. With the exception of VIO and scene reconstruction, the computation for all other components is independent of the input; therefore, the specific choice of the input dataset is not pertinent. Thus, any (monochrome) eye image for eye tracking, any frame for reprojection and hologram, and any sound sample for audio encoding and playback will produce the same result for power and performance standalone. In our experiments, we used the following datasets for these components:
OpenEDS~\cite{openeds} for eye tracking, 2560$\times$1440 pixel frames from VR Museum of Fine Art~\cite{vrMuseum} for timewarp and hologram, and 48 KHz audio clips~\cite{lectureSample,radioSample} from Freesound~\cite{freesound} for audio encoding and playback. For VIO and scene reconstruction, many off-the-shelf datasets are available~\cite{burri2016eurocDataset,HandaWhelan2014,SturmMagnenat2011,PireMujica2019}. We used Vicon Room 1 Easy, Medium, and Difficult~\cite{burri2016eurocDataset} for VIO. These sequences present varying levels of pose estimation difficulty to the VIO algorithm by varying both camera motion and level of detail in the frames. We used dyson\_lab~\cite{ElasticFusion} for scene reconstruction as it was provided by the authors of ElasticFusion.

\subsection{Metrics}
\label{subsec:implementation-metrics}

\noindent
{\bf Execution time:}
    We used NSight Systems~\cite{NSightSys} to obtain the overall execution timeline of the components and \NAME{}, including serial CPU, parallel CPU, GPU compute, and GPU graphics phases.
    VTune~\cite{VTune} (desktop only) and \texttt{perf}~\cite{perf} provided CPU hotspot analysis and hardware performance counter information.
    NSight Compute~\cite{NSightCompute} and Graphics~\cite{NSightGraphics} provided detailed information on CUDA kernels and GLSL shaders, respectively.
    We also developed a logging framework that allows \NAME{} to easily collect the wall time and CPU time of each of its components. We used this framework to obtain execution time without profiling overhead.
    We determined the contribution of a given component towards the CPU by computing the total CPU cycles consumed by that component as a fraction of the cycles used by all components.

\noindent
{\bf Power and energy:}
    For the desktop, we measured CPU power using \texttt{perf} and GPU power using \texttt{nvidia-smi}~\cite{NVIDIASMI}. On the embedded platform, we collected power and energy using a custom profiler, similar to~\cite{LengChen2019}, that monitored power rails~\cite{TegraINA} to calculate both the average power and average energy for different components of the system: CPU, GPU, DDR (DRAM), SoC (on-chip microcontrollers), and Sys (display, storage, I/O).
    

\noindent
{\bf Motion-to-photon latency:}
    We compute motion-to-photon latency as the age of the reprojected image's pose when the pixels started to appear on screen. This latency is the sum of how old the IMU sample used for pose calculation was, the time taken by reprojection itself, and the wait time until the pixels started (but not finished) to get displayed. Mathematically, this can be formulated as: $latency = t_{imu\_age} + t_{reprojection} + t_{swap}$. We do not include $t_{display}$, the time taken for all the pixels to appear on screen, in this calculation. This calculation is performed and logged by the reprojection component every time it runs. If reprojection misses vsync, the additional latency is captured in $t_{swap}$.
    
\noindent
{\bf Image Quality:}
    To compute image quality, we compare the outputs of an actual and idealized XR system. We use the Vicon Room 1 Medium dataset, which provides ground-truth poses in addition to IMU and camera data. We feed the IMU and camera data into the actual system, while feeding in the equivalent ground-truth to the idealized system. In each system, we record each image submitted by the application as well as its timestamp and pose. Due to the fact that the coordinate systems of OpenVINS and the ground truth dataset are different, we perform a trajectory alignment step to obtain aligned ground-truth poses~\cite{ZhangScaramuzza2018}. We then feed the aligned poses back into our system to obtain the correct ground-truth images.
    
    Once the images have been collected, we manually pick the starting points of the actual and idealized images sequences so that both sequences begin from the same moment in time. Then, we compute the actual and idealized reprojected images using the collected poses and source images. The reprojected images are the ones that would have been displayed to the user. If no image is available for a timestamp due to a dropped frame, we use the previous image. Finally, we compare the actual and idealized warped images to compute both SSIM and FLIP. We report 1-FLIP in this paper in order to be consistent with SSIM (0 being no similarity and 1 being identical images).
    
\noindent
{\bf Pose Error:}
    We use the data collection mechanism described above to collect actual and ground-truth poses. We calculate Absolute Trajectory Error (ATE)~\cite{SturmMagnenat2011} and Relative Pose Error (RPE)~\cite{kummerle2009measuring} using OpenVINS' error evaluation tool~\cite{OpenVINS}.
\section{Results}
\label{sec:results}

\begin{figure}[!t]
    \subfloat[Desktop]{
    \includegraphics[width=\linewidth]{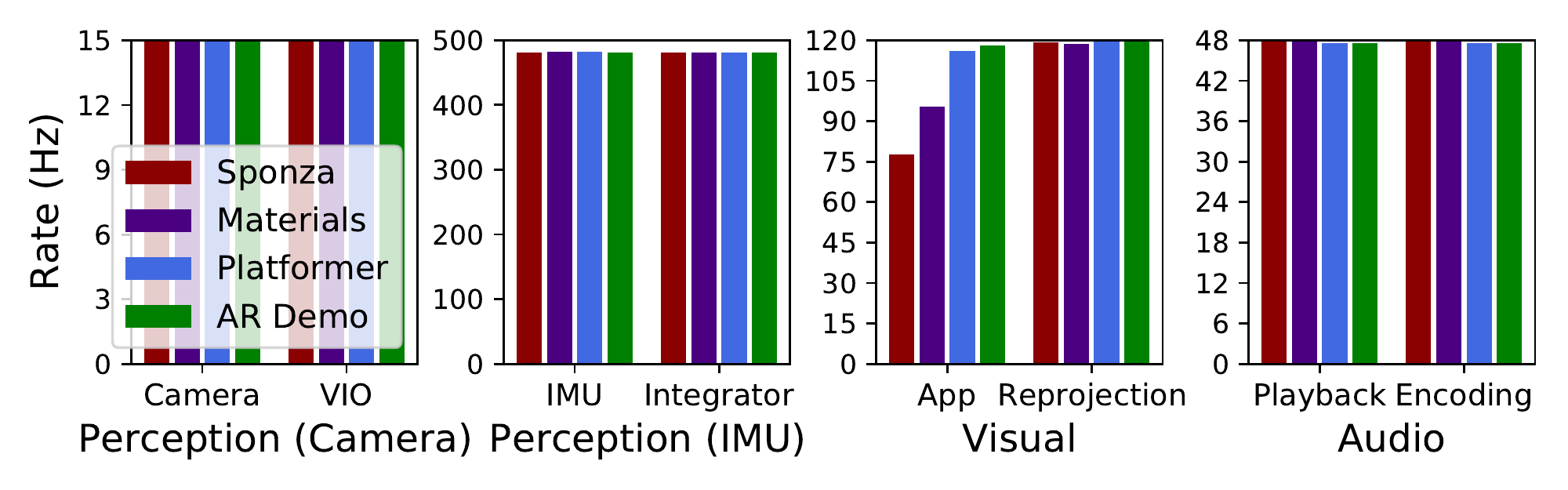}
    \label{fig:fps-desktop}
    }
    
    \subfloat[Jetson-HP]{
    \includegraphics[width=\linewidth]{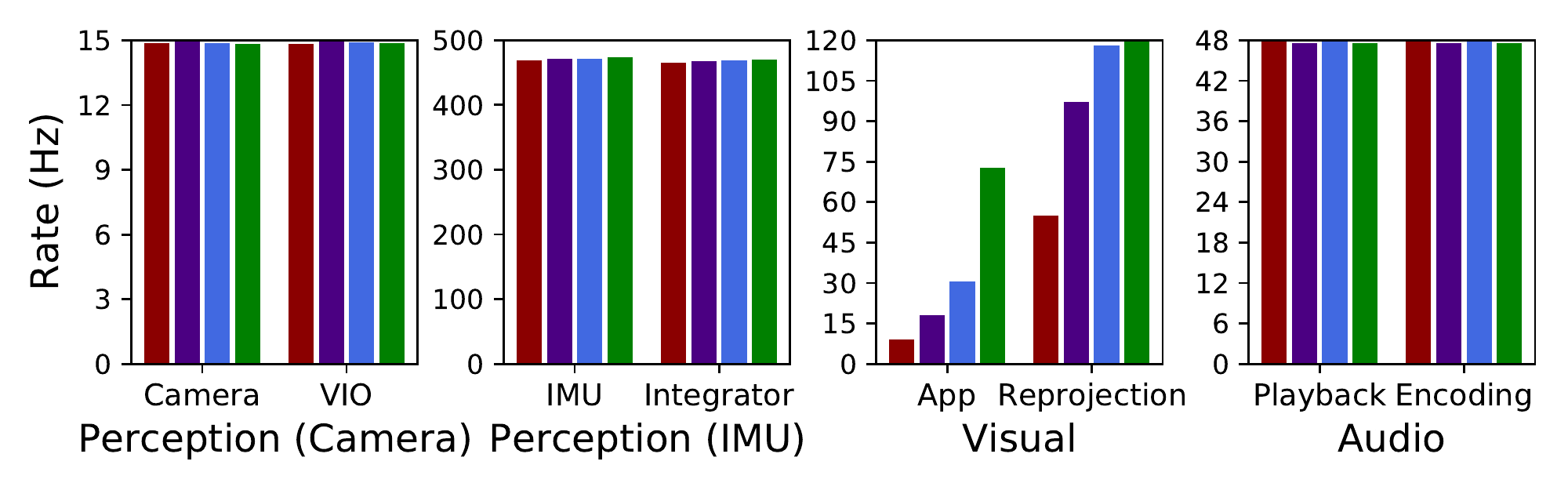}
    \label{fig:fps-jhp}
    }
    
    \subfloat[Jetson-LP]{
    \includegraphics[width=\linewidth]{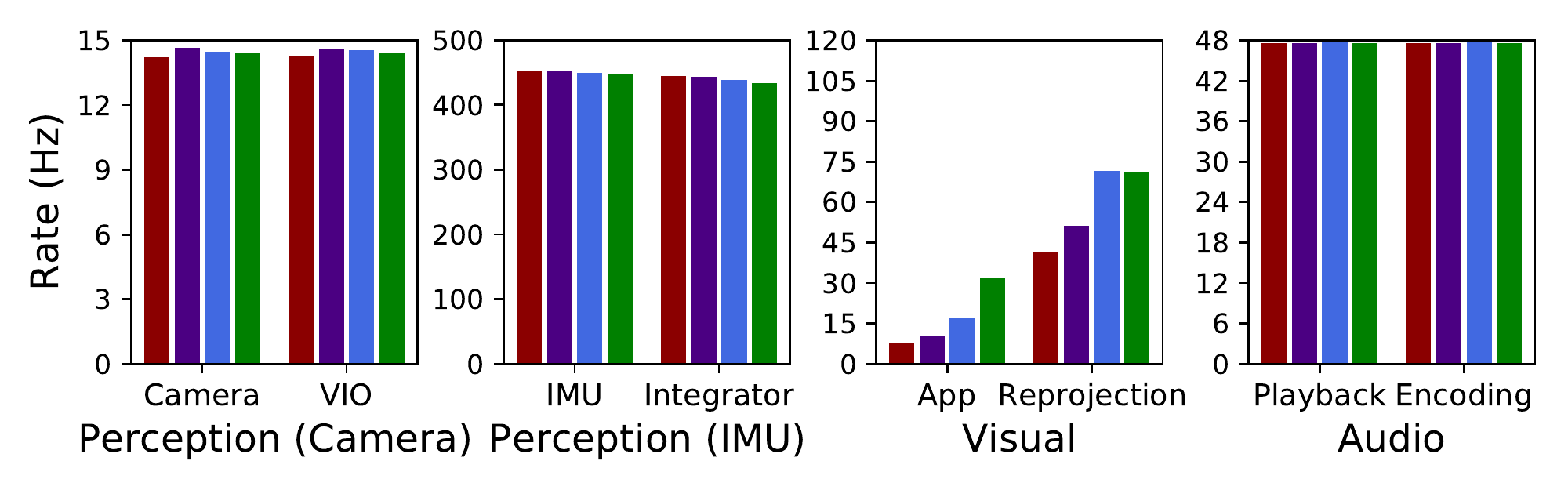}
    \label{fig:fps-jlp}
    }
    \caption{Average frame rate for each component in the different pipelines on each application and hardware platform. The y-axis is capped at the target frame rate of the pipeline.
    }
    \label{fig:system-fps}
\end{figure}

\begin{figure}[!t]
    \subfloat[Desktop]{
    \includegraphics[width=\linewidth]{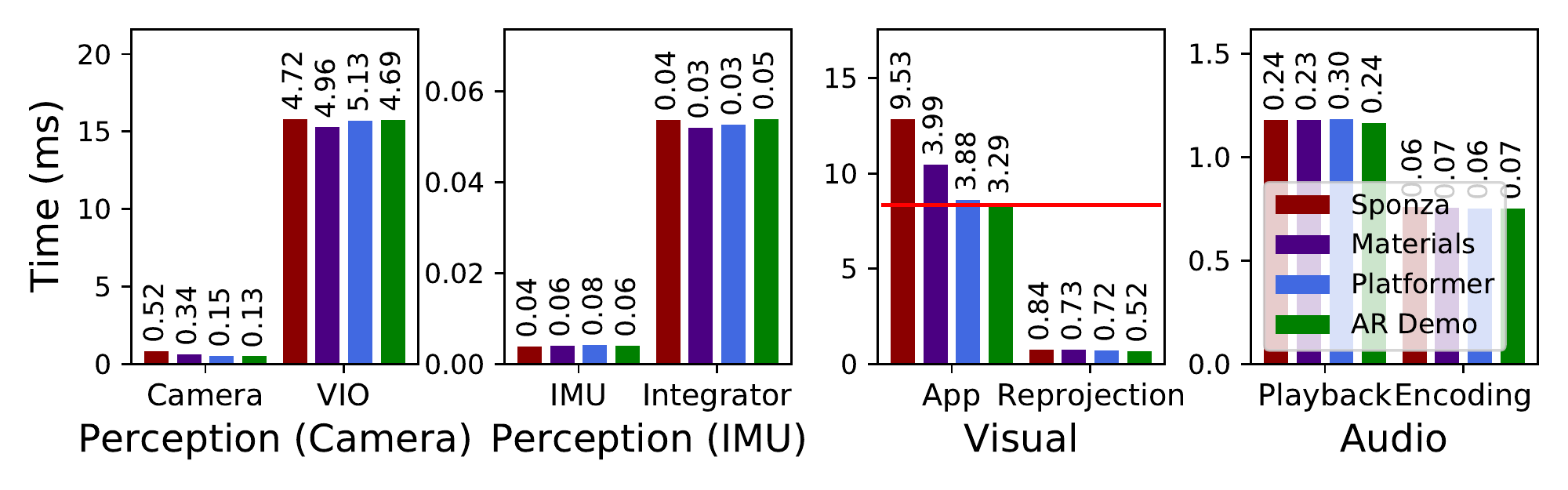}
    \label{fig:time-desktop}
    }
    
    \subfloat[Jetson-HP]{
    \includegraphics[width=\linewidth]{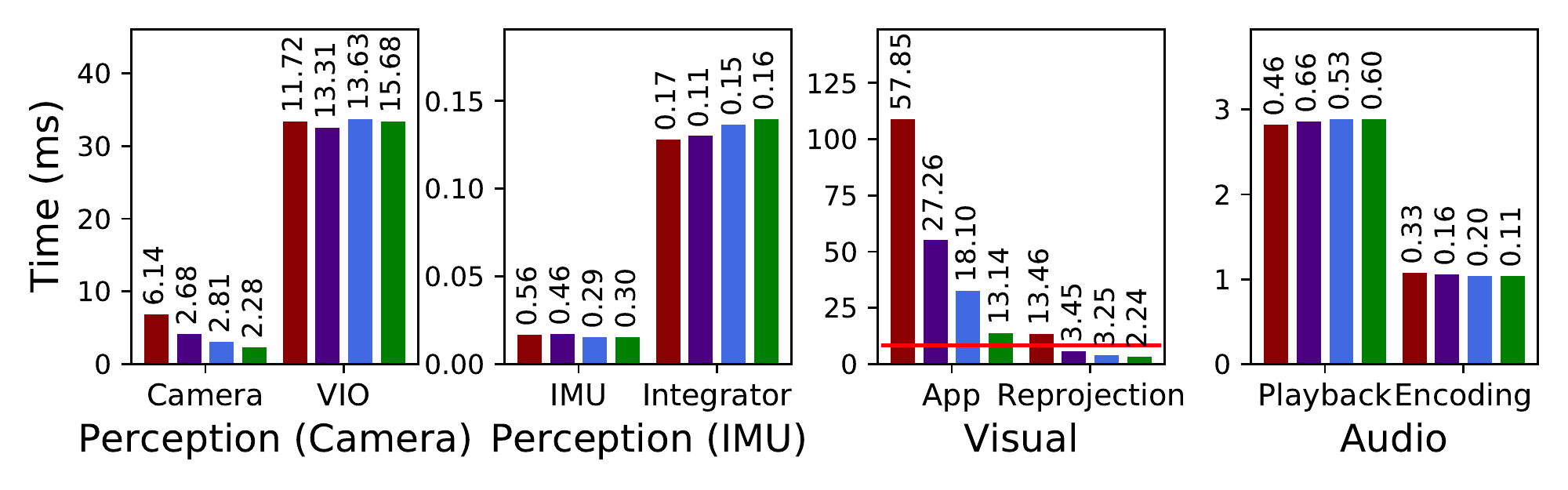}
    \label{fig:time-jhp}
    }
    
    \subfloat[Jetson-LP]{
    \includegraphics[width=\linewidth]{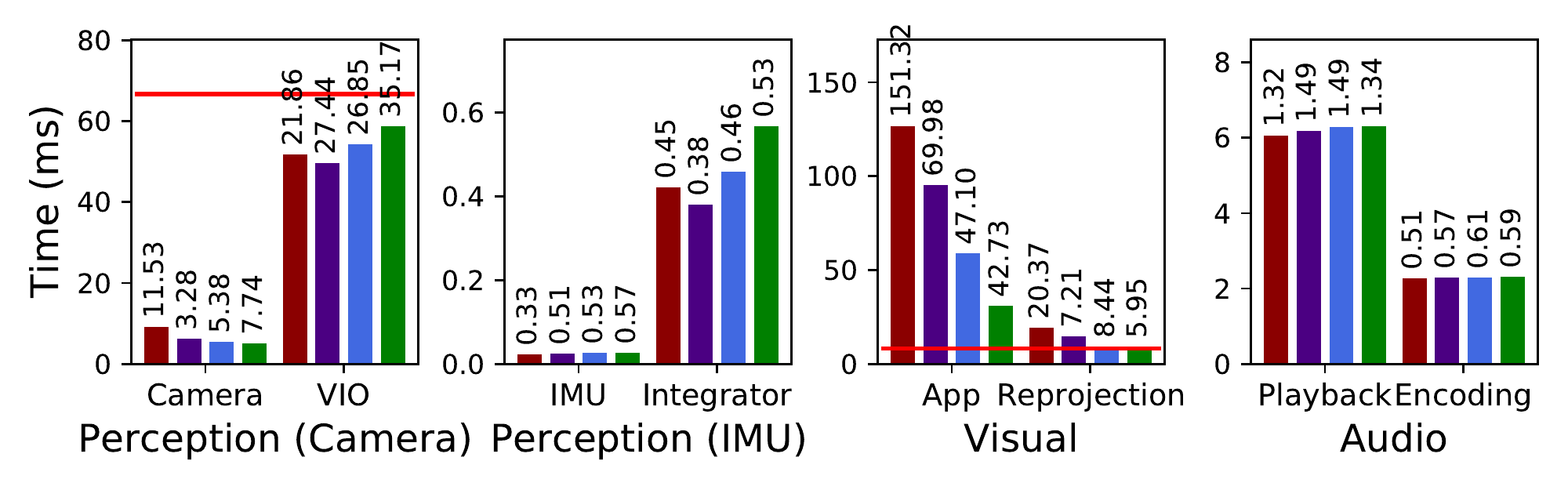}
    \label{fig:time-jlp}
    }
    \caption{Average per-frame execution time for each component for each application and hardware platform. The number on top of each bar is the standard deviation across all frames. The red horizontal line shows the maximum execution time (deadline) to meet the target frame rate. The line is not visible in graphs where the achieved time is significantly below the deadline.
    }
    \label{fig:system-time}
\end{figure}

\begin{figure}[ht]
    \centering
    \includegraphics[width=\columnwidth]{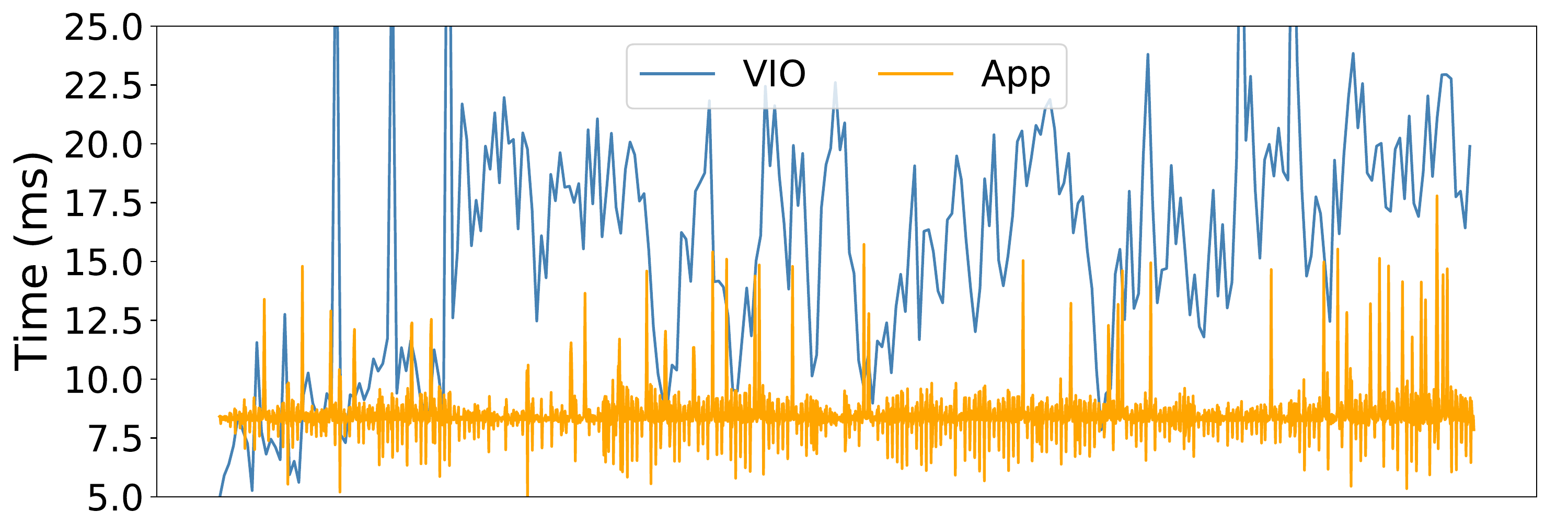}
    \includegraphics[width=\columnwidth]{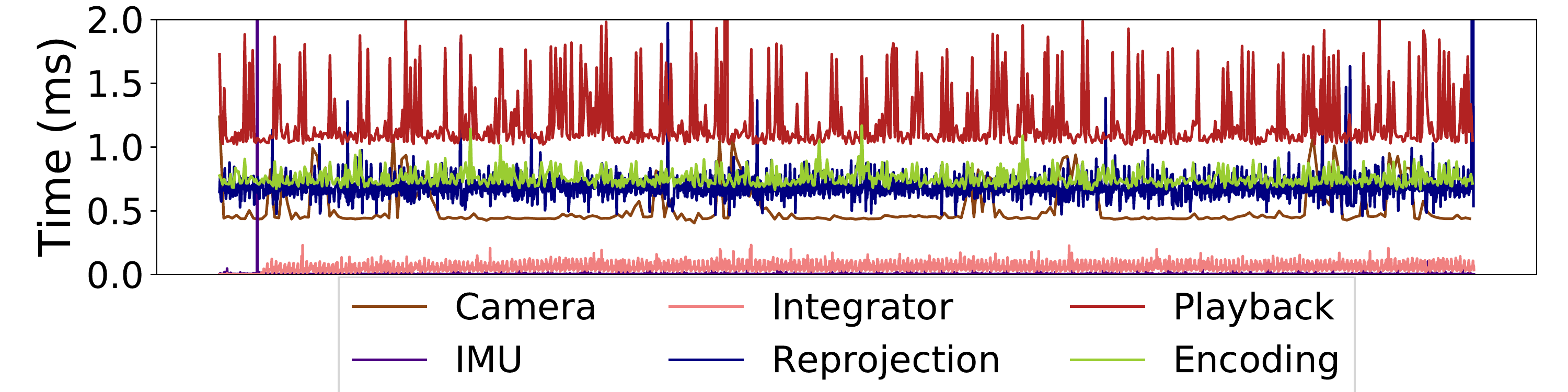}
    \caption{Per-frame execution times for Platformer on desktop. The top graph shows VIO and the application, and the bottom graph shows the remaining components. Note the different scales of the y-axes.}
    \label{fig:timeseries-sponza}
\end{figure}

\begin{figure}[ht]
    \centering
    \includegraphics[width=\linewidth]{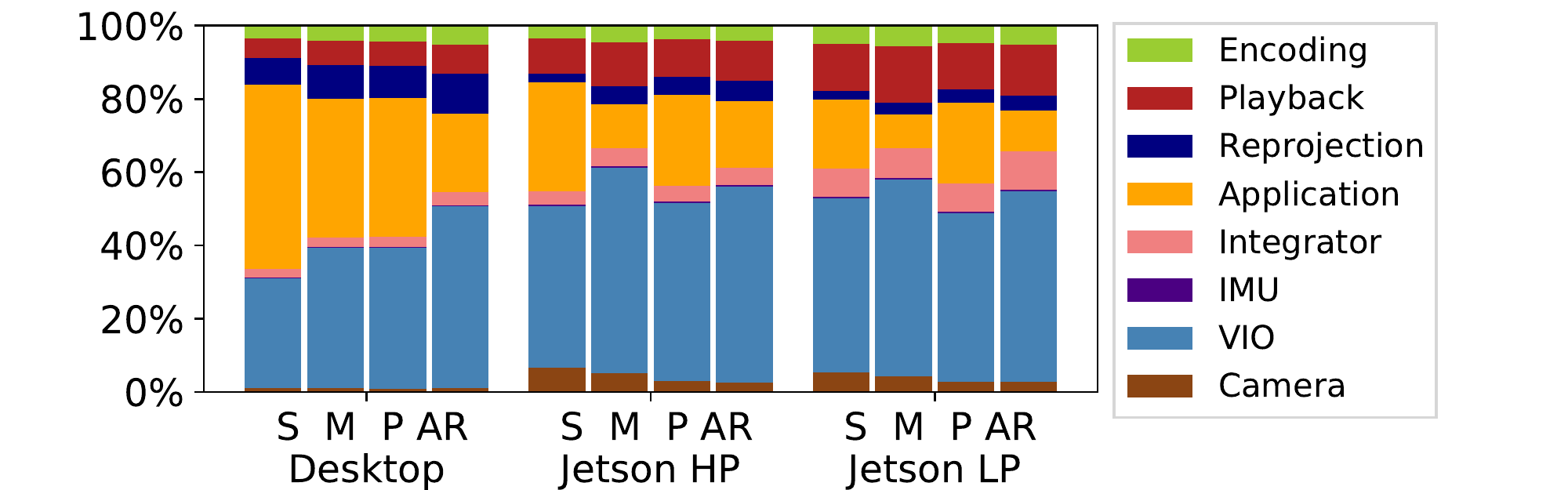}
    \caption{Contributions of \NAME{} components to CPU time.}
    \label{fig:time-breakdown}
\end{figure}




\subsection{Integrated \NAME{} System}
\label{subsec:results-integrated}

Section~\ref{subsubsec:results-performance}--\ref{subsubsec:results-qoe} respectively present system-level results for performance (execution time and throughput), power, and QoE for the integrated \NAME{} system configuration.

\subsubsection{Performance}
\label{subsubsec:results-performance}

\NAME{} consists of multiple interacting streaming pipelines and there is no single number that can capture the entirety of its performance (execution time or throughput). We present 
the performance results for \NAME{} in terms of achieved frame rate and per-frame execution time (mean and standard deviation) for each component, and compare them against the target frame rate and corresponding deadline respectively (Table~\ref{table:parameters}). 
Further, to understand the importance of each component to execution time resources, we also present the contribution of the different components to the total CPU execution cycles.

\textbf{Component frame rates.}
Figures~\ref{fig:system-fps}(a)-(c) show each component's average frame rate for each application, for a given hardware configuration. Components with the same target frame rate are presented in the same graph, with the maximum value on the y-axis representing this target frame rate. Thus, we show separate graphs for components in the perception, visual, and audio pipelines, with the perception pipeline further divided into two graphs representing the camera and IMU driven components.

Focusing on the desktop, Figure~\ref{fig:fps-desktop} shows that virtually all components meet, or almost meet, their target frame rates (the application component for Sponza and Materials are the only exceptions). The IMU components are slightly lower than the target frame rate only because of scheduling non-determinism at the 2 ms period required by these components. This high performance, however, comes with a significant power cost.

Moving to the lower power Jetson, we find more components missing their target frame rates. With Jetson-LP, only the audio pipeline is able to meet its target. The visual pipeline components  -- application and reprojection -- are both severely degraded in all cases for Jetson-LP and most cases for Jetson-HP, sometimes by more than an order of magnitude. (Recall that the application and reprojection missing vsync results in the loss of an entire frame. In contrast,  other components can catch up on the next frame even after missing the deadline on the current frame, exhibiting higher effective frame rates.) 

Although we assume modern display resolutions and refresh rates, future systems will support larger and faster displays with larger field-of-view and will integrate more components, further stressing the entire system. 

\textbf{Execution time per frame.}
Figures~\ref{fig:system-time}(a)-(c) show the average execution time (wall clock time) per frame for a given component, organized similar to Figure~\ref{fig:system-fps}. The horizontal line on each graph shows the target execution time or deadline, which is the reciprocal of the target frame rate. 
Note that the achieved frame rate in Figure~\ref{fig:system-fps} cannot exceed the target frame rate as it is controlled by the runtime.  The execution time per frame, however, can be arbitrarily low or high (because we don't preempt an execution that misses its deadline). The number at the top of each bar shows the standard deviation across all frames. 
 
The mean execution times follow trends similar to those for frame rates. The standard deviations for execution time are surprisingly significant in many cases. For a more detailed view, Figure~\ref{fig:timeseries-sponza} shows the execution time of each frame of each \NAME{} component during the execution of Platformer on the desktop (other timelines are omitted for space). The components are split across two graphs for clarity. We expect variability in VIO (blue) and application (yellow) since these computations are known to be input-dependent. However, we see significant variability in the other components as well. This variability is due to scheduling and resource contention,
motivating the need for better resource allocation, partitioning, and scheduling. A consequence of the variability is that although the mean per-frame execution time for some components is comfortably within the target deadline, several frames do miss their deadlines; e.g., VIO in Jetson-LP. Whether this affects the user's QoE depends on how well the rest of the system compensates for these missed deadlines. For example, even if the VIO runs a little behind the camera, the IMU part of the perception pipeline may be able to compensate. Similarly, if the application misses some deadlines, reprojection may be able to compensate.  Section~\ref{subsubsec:results-qoe} provides results for system-level QoE metrics that address these questions.
 
\textbf{Distribution of cycles.} Figure~\ref{fig:time-breakdown} shows the relative attribution of the total cycles consumed in the CPU to the different \NAME{} components for the different applications and hardware platforms. These figures do not indicate wall clock time since one elapsed cycle with two busy concurrent threads will contribute two cycles in these figures. Thus, these figures indicate total CPU resources used. We do not show GPU cycle distribution because the GPU timers significantly perturbed the rest of the execution -- most of the GPU cycles are consumed by the application with the rest used by reprojection.

Focusing first on the desktop, Figure~\ref{fig:time-breakdown} shows that VIO and the application are the largest contributors to CPU cycles, with one or the other dominating, depending on the application. Reprojection and audio playback follow next, becoming larger relative contributors as the application complexity reduces. Although reprojection does not exceed 10\% of the total cycles, it is a dominant contributor to motion-to-photon latency and, as discussed below, cannot be neglected for optimization. 

Jetson-HP and Jetson-LP show similar trends except that we find that the application's and reprojection's relative contribution decreases while other components such as the IMU integrator become more significant relative to the desktop.
This phenomenon occurs because with the more resource-contrained Jetson configurations, the application and reprojection often miss their deadline and are forced to skip the next frame (Section~\ref{subsec:runtime}). Thus, the overall work performed by these components reduces, but shows up as poorer end-to-end QoE metrics discussed later (Section~\ref{subsubsec:results-qoe}).

\begin{figure}[ht]
    \subfloat[Total Power (W)]{
    \includegraphics[width=0.5\linewidth]{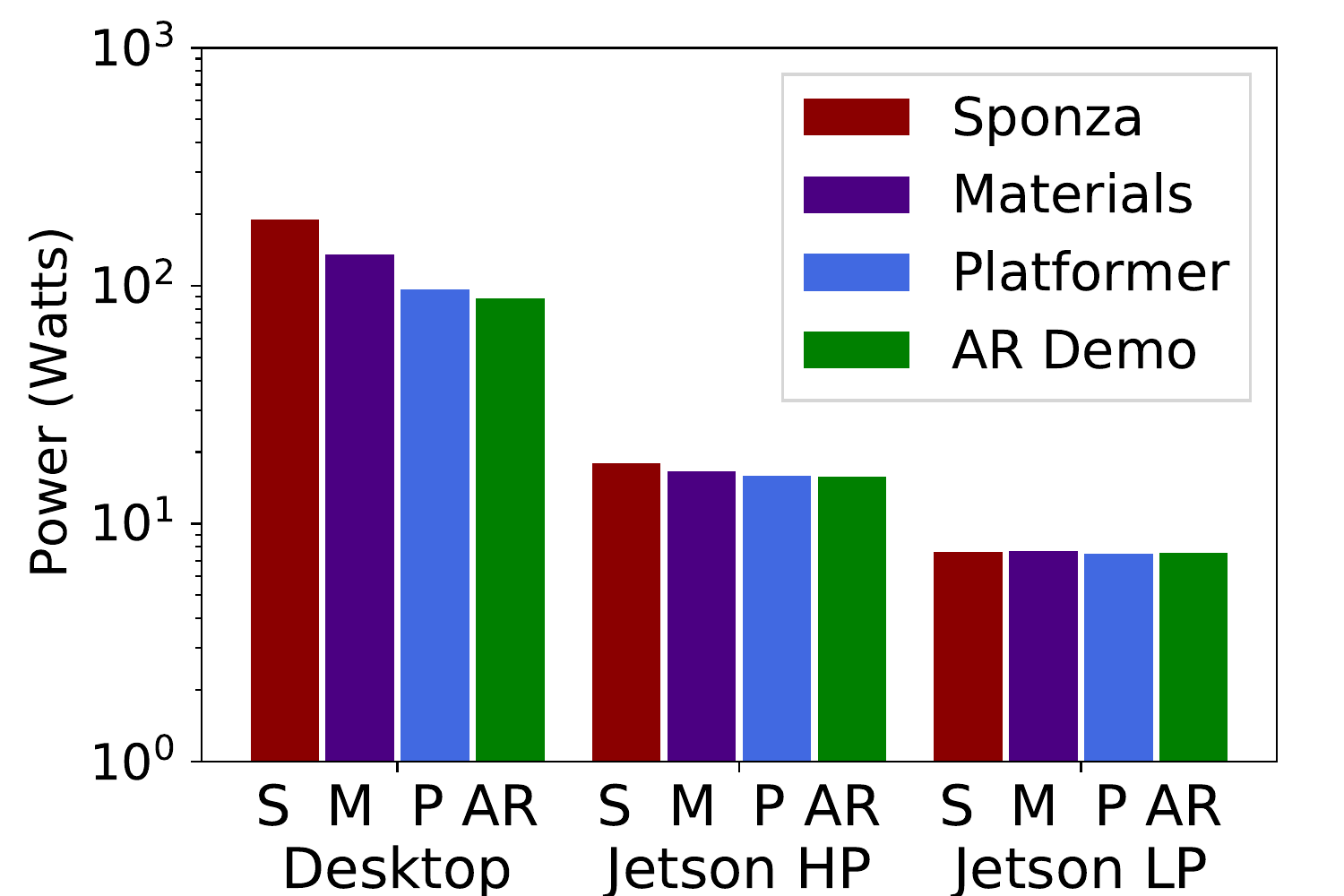}
    \label{fig:total-power}
    }
    \subfloat[Power Breakdown]{
    \includegraphics[width=0.5\linewidth]{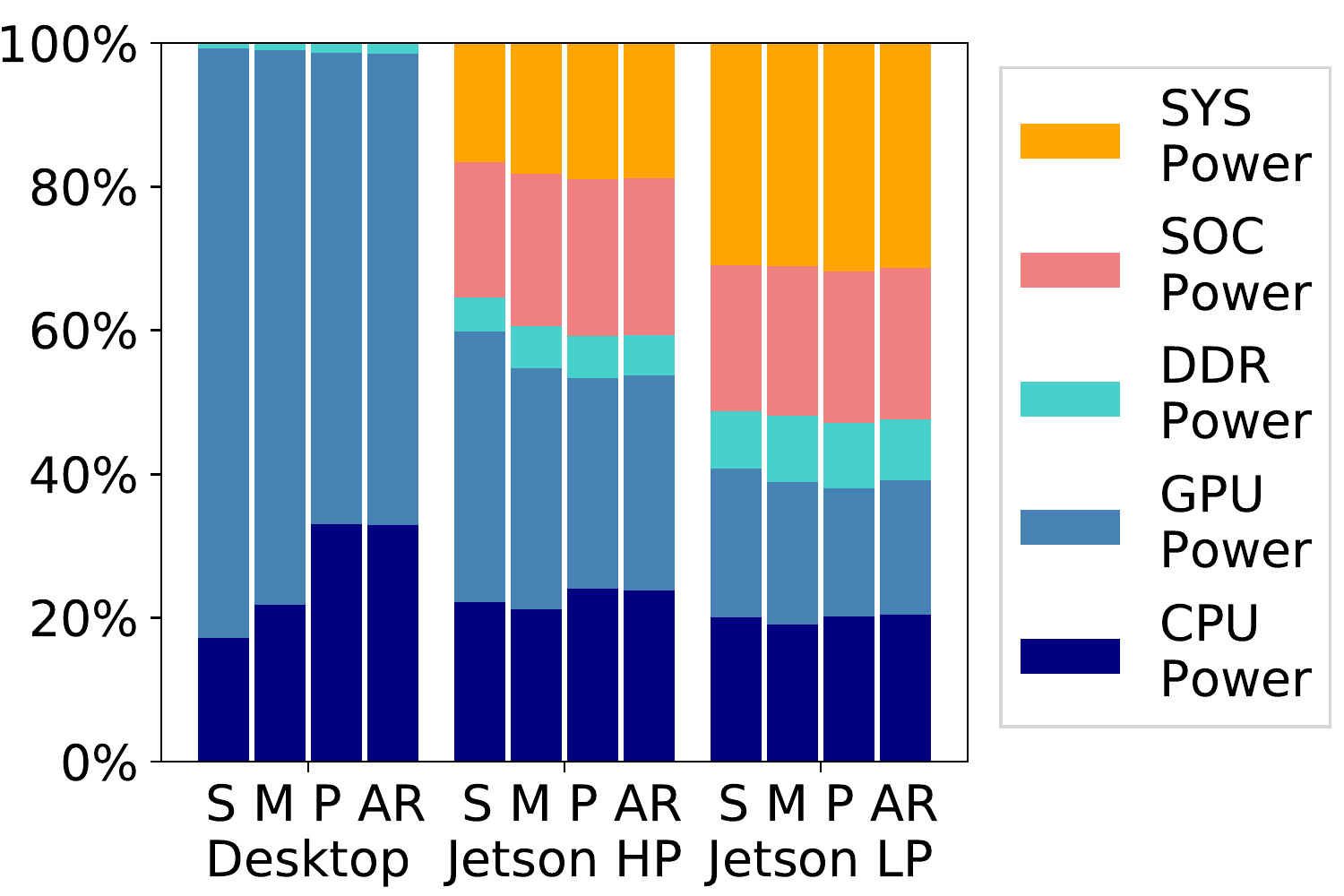}
    \label{fig:power-breakdown}
    }
    \caption{(a) Total power (note log scale) and (b) relative contribution to power by different hardware units for each application and hardware platform.}
    \label{fig:system-power}
\end{figure}

\subsubsection{Power}
\label{subsubsec:results-power}

\textbf{Total Power.} 
Figure~\ref{fig:total-power} shows the total power consumed by \NAME{} running each application on each hardware platform. 
The power gap from the ideal (Table~\ref{table:requirements}) is severe on all three platforms.
As shown in Figure~\ref{fig:total-power}, Jetson-LP, the lowest power platform, is two orders of magnitude off in terms of the ideal power while the desktop is off by three. As with performance, larger resolutions, frame rates, and field-of-views, and more components would further widen this gap.
    
\textbf{Contribution to power from different hardware components.}
Figure~\ref{fig:power-breakdown} shows the relative contribution of different hardware units to the total power, broken down as CPU, GPU, DDR, SoC , and Sys. These results clearly highlight the need for studying all aspects of the system. Although the GPU dominates power on the desktop, that is not the case on Jetson. SoC and Sys power are often ignored, but doing so does not capture the true power consumption of XR workloads, which typically exercise all the aforementioned hardware units -- SoC and Sys consume more than 50\% of total power on Jetson-LP. Thus, reducing the power gap for XR components would require optimizing system-level hardware components as well.

\subsubsection{Quality of Experience}
\label{subsubsec:results-qoe}

\begin{table}[t!]
\centering
{
    \caption{
        Motion-to-photon latency in milliseconds (mean$\pm$std dev), without $t_{display}$. Target is 20 ms for VR and 5 ms for AR (Table~\ref{table:requirements}).
    }
    \scriptsize
    \begin{tabular}{c c c c}
    \textbf{Application}    & \textbf{Desktop}  & \textbf{Jetson-hp}    & \textbf{Jetson-lp} \\ \toprule
    Sponza                  & $3.1 \pm 1.1$     & $13.5 \pm 10.7$       & $19.3 \pm 14.5$ \\ \hline
    Materials               & $3.1 \pm 1.0$     & $7.7 \pm 2.7$         & $16.4 \pm 4.9$ \\ \hline
    Platformer              & $3.0 \pm 0.9$     & $6.0 \pm 1.9$         & $11.3 \pm 4.7$ \\ \hline
    AR Demo                 & $3.0 \pm 0.9$     & $5.6 \pm 1.4$         & $12.0 \pm 3.4$
    
    \end{tabular}
    \label{table:results-mtp}
}
\end{table}

\begin{figure}[ht]
    \centering
    \includegraphics[width=\linewidth]{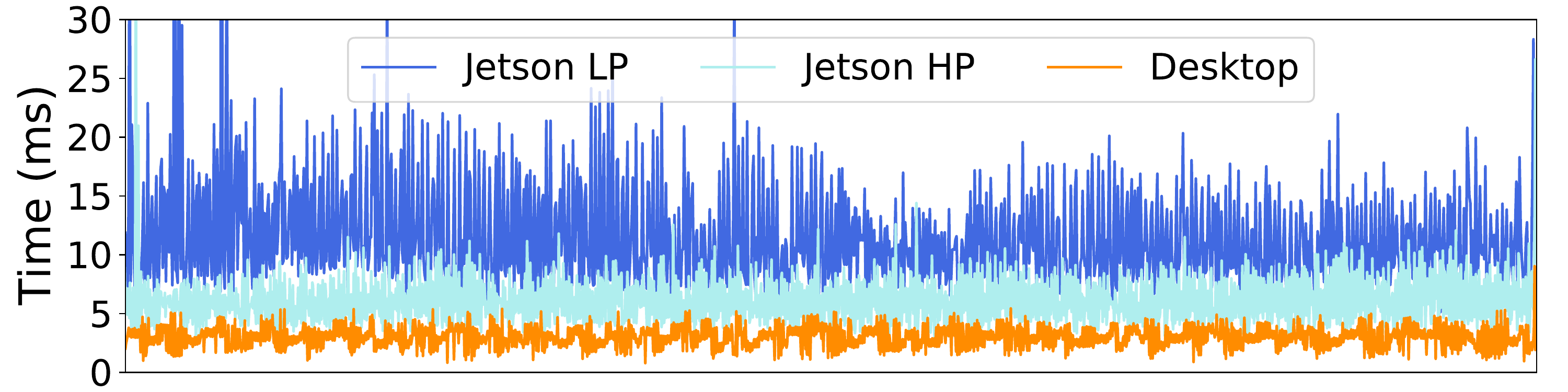}
    \caption{Motion-to-photon latency per frame of Platformer.
    }
\label{fig:mtp-platformer}
\end{figure}

Quality of an XR experience is often determined through user studies~\cite{bauman2017towards,ArifinSastria2018,zhu2020better}; however, these can be expensive, time consuming, and subjective. \NAME{}, therefore, provides several quantitative metrics to measure QoE. We report both results of visual examination and quantitative metrics below.

\textbf{Visual examination.} A detailed user study is outside the scope of this work, but there were several artifacts in the displayed image that were clearly visible. As indicated by the performance metrics, the desktop displayed smooth images for all four applications. Jetson-HP showed perceptibly increased judder for Sponza. Jetson-LP showed dramatic pose drift and clearly unacceptable images for Sponza and Materials, though it was somewhat acceptable for the less intense Platformer and AR Demo. As discussed in Section~\ref{subsubsec:results-performance}, the average VIO frame rate for Jetson-LP stayed high, but the variability in the per-frame execution time resulted in many missed deadlines, which could not be fully compensated by the IMU or reprojection. These effects are quantified with the metrics below.

\textbf{Motion-to-photon latency (MTP).}
Table~\ref{table:results-mtp} shows the mean and standard deviation of MTP for all cases. Figure~\ref{fig:mtp-platformer} shows MTP for each frame over the execution of Platformer on all hardware (we omit the other applications for space).
Recall the target MTP is 20 ms for VR and 5 ms for AR (Table~\ref{table:requirements}) and that our measured numbers reported do not include $t_{display}$ (4.2 ms).

Table~\ref{table:results-mtp} and our detailed per-frame data shows that the desktop can achieve the target VR MTP for virtually all frames. For AR, most cases appear to meet the target, but adding $t_{display}$ significantly exceeds the target.
Both Jetsons cannot make the target AR MTP for an average frame. Jetson-HP is able to make the VR target MTP for the average frame for all applications and for most frames for all except Sponza (even after adding $t_{display}$). Jetson-LP shows a significant MTP degradation -- on average, it still meets the target VR MTP, but both the mean and variability increase with increasing complexity of the application until Sponza is practically unusable (19.3 ms average without $t_{display}$ and 14.5 ms standard deviation).
Thus, the MTP data collected by \NAME{} is consistent with the visual observations reported above.

\begin{table}[t!]
\centering
{
    \caption{
        Image and pose quality metrics (mean$\pm$std dev) for Sponza.
    }
    \scriptsize
    \begin{tabular}{c c c c c}
    \textbf{Platform}   & \textbf{SSIM}     & \textbf{1-FLIP}     & \textbf{ATE/degree}  & \textbf{ATE/meters} \\ \toprule
    Desktop             & $0.83 \pm 0.04$   & $0.86 \pm 0.05$         & $8.6 \pm 6.2$   & $0.33 \pm 0.15$ \\ \hline
    Jetson-hp           & $0.80 \pm 0.05$   & $0.85 \pm 0.05$         & $18  \pm  13$   & $0.70 \pm 0.33$ \\ \hline
    Jetson-lp           & $0.68 \pm 0.09$   & $0.65 \pm 0.17$         & $138 \pm  26$   & $13   \pm   10$ \\    
    \end{tabular}
    \label{table:results-offline-metrics}
}
\end{table}

\textbf{Offline metrics for image and pose quality.} 
Table~\ref{table:results-offline-metrics} shows the mean and standard deviation for SSIM, 1-FLIP, ATE/degree, and ATE/distance (Section~\ref{subsec:metrics} and~\ref{subsec:implementation-metrics}) for Sponza on all hardware configurations. (We omit other applications for space.) Recall that these metrics require offline analysis and use a different offline trajectory dataset for VIO (that comes with ground truth) from the live camera trajectory reported in the rest of this section. Nevertheless, the visual experience of the applications is similar. We find that all metrics degrade as the hardware platform becomes more constrained. While the trajectory errors reported are clearly consistent with the visual experience, the SSIM and FLIP values seem deceptively high for the Jetsons (specifically the Jetson-LP where VIO shows a dramatic drift). Quantifying image quality for XR is known to be challenging~\cite{ZhangZhao2017, KimLim2019}. While some work has proposed the use of more conventional graphics-inspired metrics such as SSIM and FLIP, this is still a topic of ongoing research~\cite{YangLiu2018, LimKim2018}. \NAME{} is able to capture the proposed metrics, but our work also motivates and enables research on better image quality metrics for XR experiences.

\subsubsection{Key Implications for Architects}

Our results characterize end-to-end performance, power, and QoE of an XR device, exposing new research opportunities for architects and demonstrating \NAME{} as a unique testbed to enable exploration of these opportunities, as follows. 

{\bf Performance, power, and QoE gaps.} Our results quantitatively show that collectively there is several orders of magnitude performance, power, and QoE gap between current representative desktop and embedded class systems and the goals in Table~\ref{table:requirements}.
The above gap will be further exacerbated with higher fidelity displays and the addition of more components for a more feature-rich XR experience (e.g., scene reconstruction, eye tracking, hand tracking, and holography).

While the presence of these gaps itself is not a surprise, we provide the first such quantification and analysis. This provides insights for directions for architecture and systems research (below) as well as demonstrates \NAME{} as a one-of-a-kind testbed that can enable such research.

{\bf No dominant performance component.} Our more detailed results show that there is no one component that dominates all the metrics of interest.
While the application and VIO seem to dominate CPU cycles, reprojection latency is critical to MTP, and audio playback takes similar or more cycles as reprojection. Since image quality relies on accurate poses, frequent operation of the IMU integrator is essential to potentially compensate for missed VIO deadlines. As mentioned earlier, Hologram dominated the GPU and was not even included in the integrated configuration studied since it precluded meaningful results on the Jetson-LP.
Thus, to close the aforementioned gaps, \emph{all} components have to be considered together, even those that may appear relatively inexpensive at first glance.
Moreover, we expect this diversity of important components to only increase as more components are included for more feature rich experiences.
These observations coupled with the large performance-power gaps above indicate a rich space for hardware specialization research, including research in automated tool flows to determine what to accelerate, how to accelerate these complex and diverse codes, and how to best exploit accelerator-level parallelism~\cite{HillReddi}.

{\bf Full-system power contributions.}
Our results show that addressing the power gap requires considering \textit{system-level} hardware components, such as display and other I/O, which consume significant energy themselves. This also includes numerous sensors. While we do not measure individual sensor power, it is included in Sys power on Jetson, which is significant. This motivates research in unconventional architecture paradigms such as on-sensor computing to save I/O power.

{\bf Variability and scheduling.}
Our results show large variability in per-frame processing times in many cases, either due to inherent input-dependent nature of the component (e.g., VIO) or due to resource contention from other components. This variability poses challenges to, and motivates research directions in, scheduling and resource partitioning and allocation of hardware resources. As presented in Table~\ref{table:parameters}, there are a multitude of parameters that need to be tuned for optimal performance of the system.
This paper performs manual tuning to fix these parameters; however, ideally they would adapt to the performance variations over time, co-optimized with scheduling and resource allocation decisions designed to meet real-time deadlines within power constraints to provide maximal QoE.
\NAME{} provides a flexible testbed to enable such research.

{\bf End-to-end quality metrics.}
The end-to-end nature of \NAME{} allows it to provide end-to-end quality metrics such as MTP and see the impact of optimizations on the final displayed images. Our results show that per-component metrics (e.g., VIO frame rate) are insufficient to determine the impact on the final user experience (e.g., as in Jetson-LP for Sponza). At the same time, there is a continuum of acceptable experiences (as in Sponza for desktop and Jetson-HP). Techniques such as approximate computing take advantage of such applications but often focus on subsystems that make it hard to assess end user impact. \NAME{} provides a unique testbed to develop cross-layer approximate computing techniques that can be driven by end-user experience. Our results also motivate work on defining more perceptive quantitative metrics for image quality and \NAME{} provides a testbed for such research.
    
\subsection{Per Component Analysis}
\label{subsec:results-component}

\begin{figure}[ht]
    \centering
    \includegraphics[width=\linewidth]{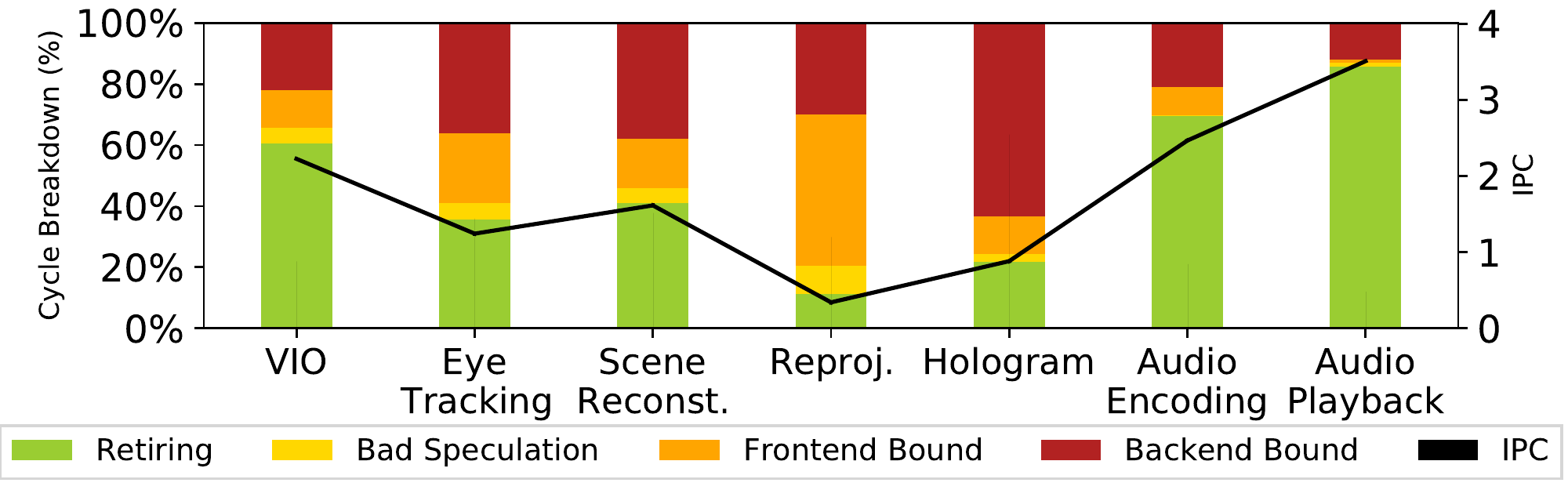}
    \caption{Cycle breakdown and IPC of \NAME{} components.
    }
\label{fig:microarchitecture}
\end{figure}

\begin{figure*}[!t]
    \centering
    \subfloat[VIO]{
    \includegraphics[width=0.135\textwidth]{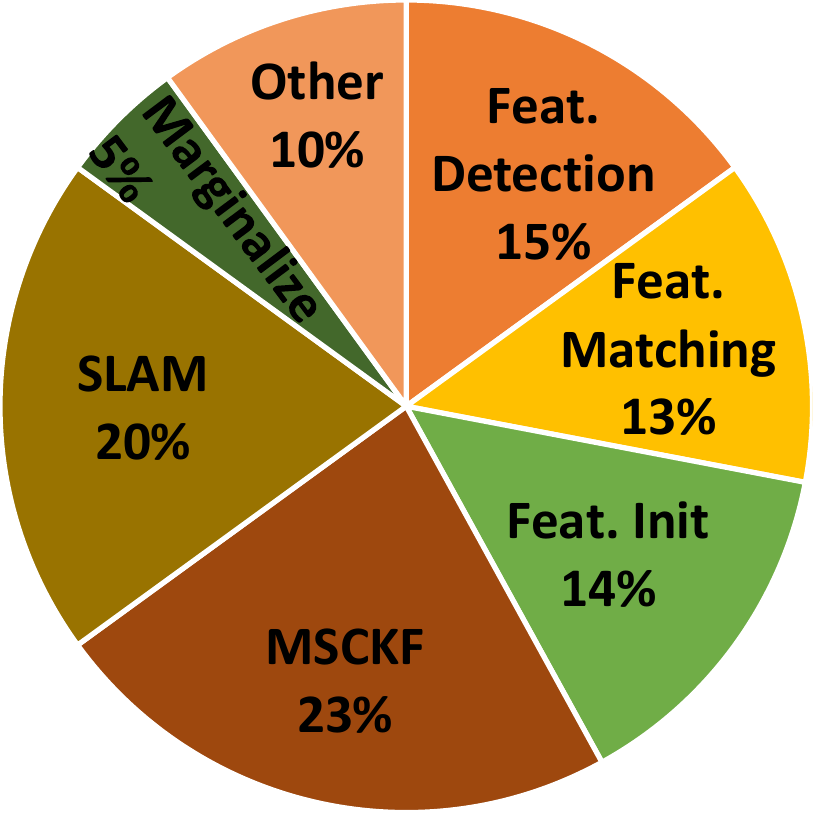}
    \label{fig:task-breakdown-vins}
    }
    \subfloat[Eye Tracking]{
    \includegraphics[width=0.135\textwidth]{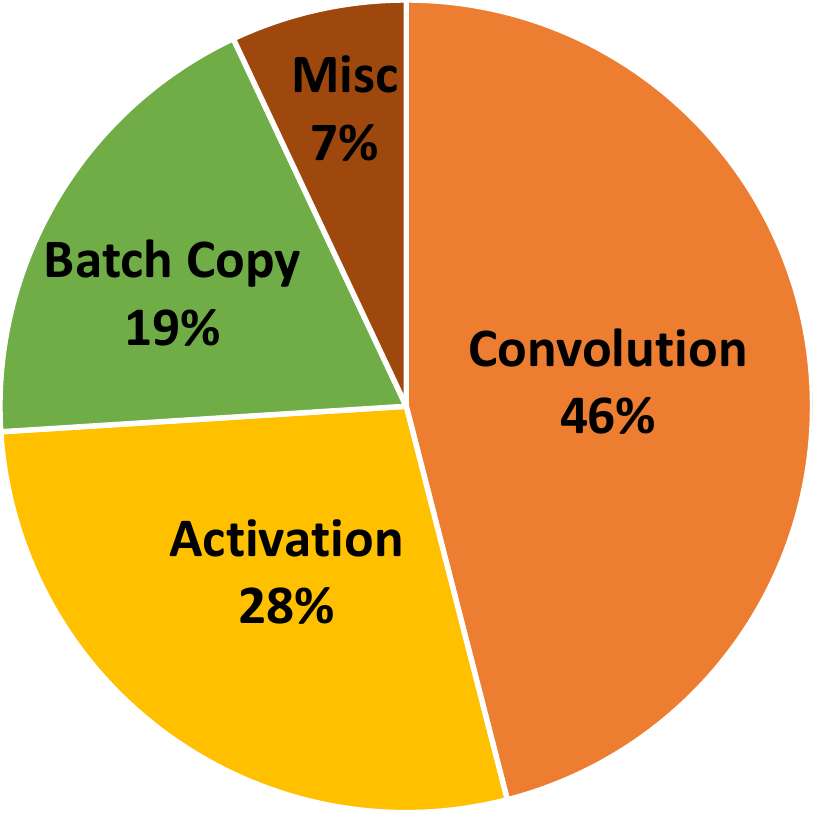}
    \label{fig:task-breakdown-ritnet}
    }
    \subfloat[Scene Reconstruction]{
    \includegraphics[width=0.135\textwidth]{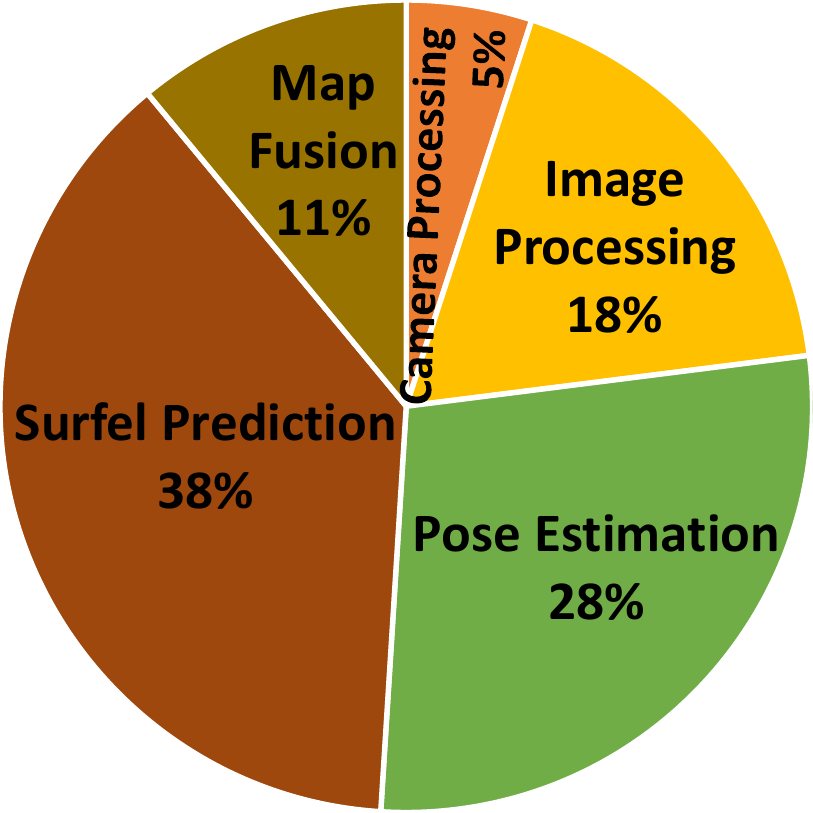}
    \label{fig:task-breakdown-ef}
    }
    \subfloat[Reprojection]{
    \includegraphics[width=0.135\textwidth]{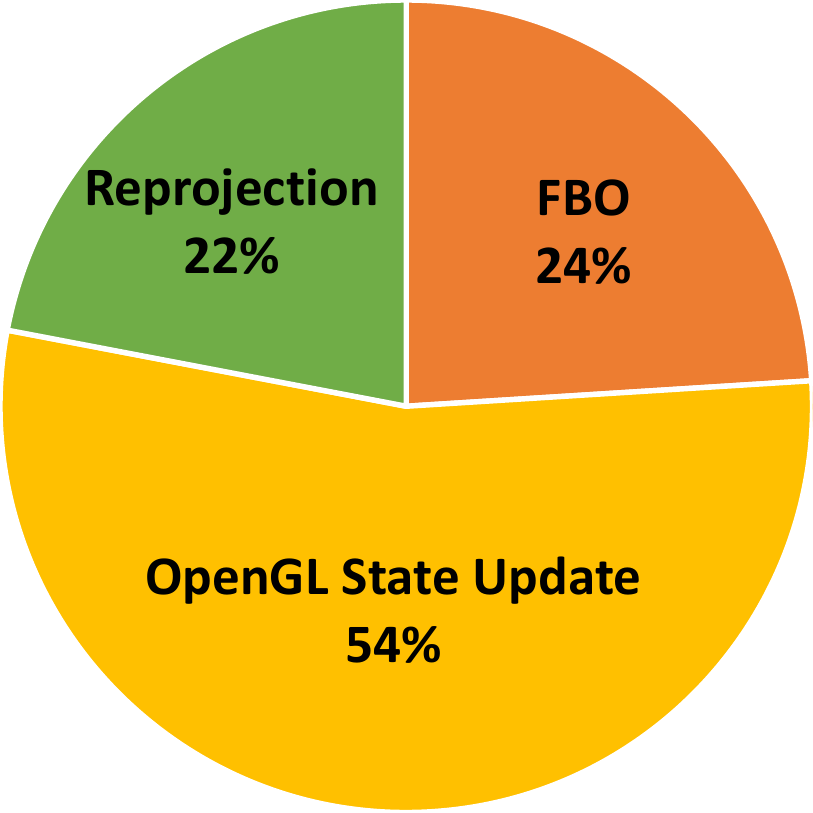}
    \label{fig:task-breakdown-tw}
    }
    \subfloat[Hologram]{
    \includegraphics[width=0.135\textwidth]{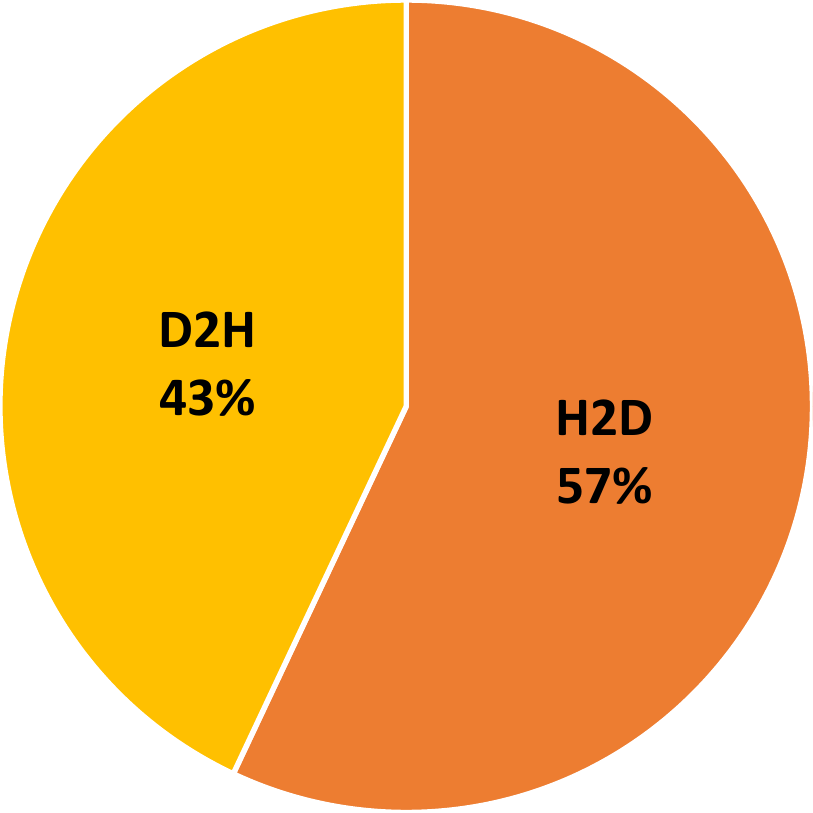}
    \label{fig:task-breakdown-hologram}
    }
    \subfloat[Audio Encoding]{
    \includegraphics[width=0.135\textwidth]{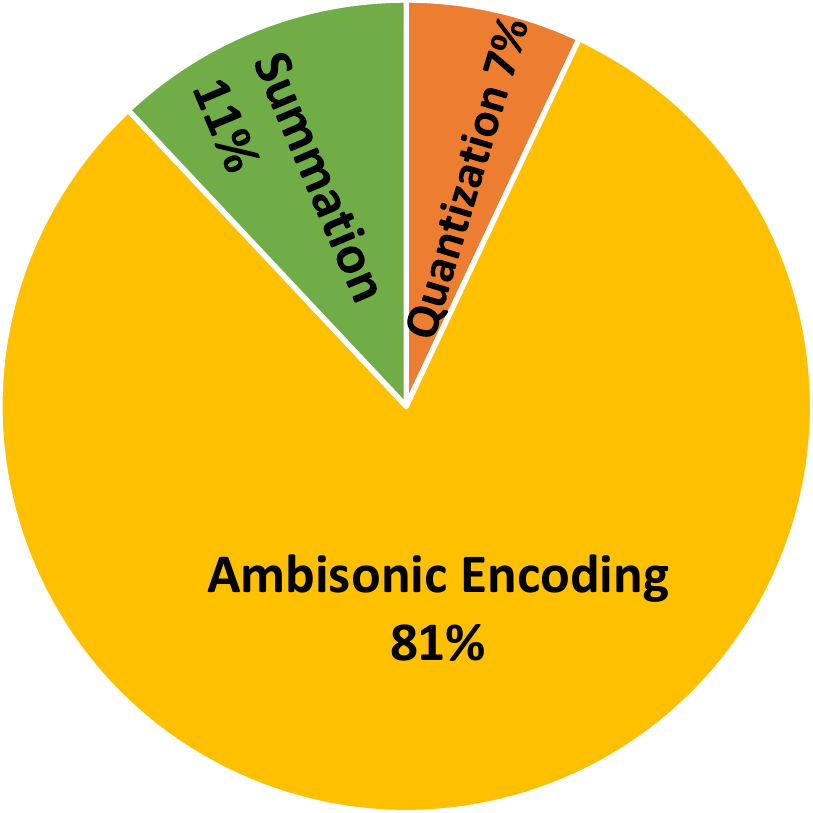}
    \label{fig:task-breakdown-record}
    }
    \subfloat[Audio Playback]{
    \includegraphics[width=0.135\textwidth]{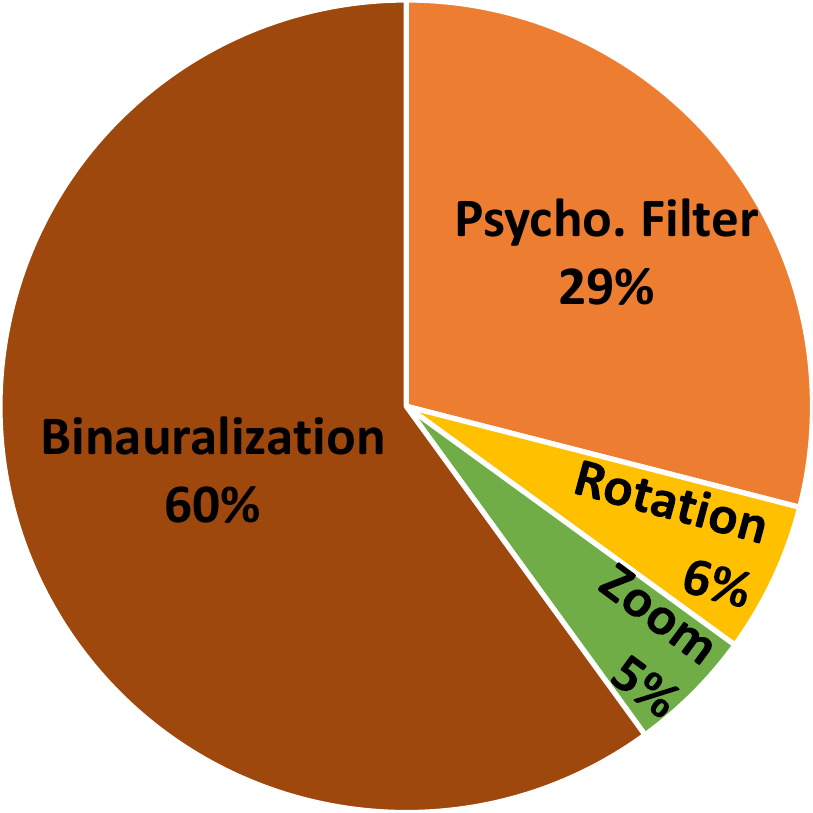}
    \label{fig:task-breakdown-playback}
    }
    \caption{Execution time breakdown into tasks of key components on the desktop. There is a large diversity in tasks and no task dominates.
    All tasks of VIO, Audio Encoding, and Audio Playback run on the CPU. The only multithreaded tasks are ``other'' and ``feature matching'' in VIO. All tasks of Eye Tracking, Scene Reconstruction, and Hologram run on the GPU. In Reprojection, ``reprojection'' runs entirely on the GPU while the other two tasks run on both the CPU and GPU.
    }
    \label{fig:task-breakdown}
\end{figure*}

The results so far motivate specialization of \NAME{} components to meet the performance-power-QoE gap as seen through the end-to-end characterization. This section examines the components in isolation to provide insights on how to specialize. We performed an extensive deep dive analysis of all the components.
Tables~\ref{table:openvins}--~\ref{table:playback} summarize the tasks within each component, including the key computation and memory patterns.
Figure~\ref{fig:task-breakdown} illustrates the contribution of the tasks to a component's execution time. Figure~\ref{fig:microarchitecture} shows the CPU IPC and cycles breakdown for each component in aggregate.
Space restrictions prevent a detailed deep dive for each component, but we present two brief case studies.

    {\bf Component Diversity}
    The components present unique characteristics. Figure~\ref{fig:microarchitecture} shows a wide range in both IPC and hardware utilization for the different components. Figure~\ref{fig:task-breakdown} shows that the tasks are typically unique to a component (Section~\ref{sec:pipeline}).

    {\bf Task Diversity}
        There is a remarkable diversity of algorithmic tasks, ranging from feature detection to map fusion to signal processing and more.
        These tasks are differently amenable for execution on the CPU, GPU-compute, or GPU-graphics. 
       
      
    {\bf Task Dominance}        
        No component is composed of just one task.
        The most homogeneous is audio encoding where ambisonic encoding is 81\% of the total -- but accelerating just that would limit the overall speedup to 5.3$\times$ by Amdahl's law, which may not be sufficient given the power and performance gap shown in Section~\ref{subsec:results-integrated}.
        VIO is the most diverse, with seven major tasks and many more sub-tasks.
        
        
        
        
    {\bf Input-Dependence}
        Although we saw significant per-frame execution time variability in all components in the integrated system, the only components that exhibit such variability standalone are VIO and scene reconstruction. 
        VIO shows a range of execution time throughout its execution, with a coefficient of variation from 17\% to 26\% across the studied datasets~\cite{burri2016eurocDataset}. Scene reconstruction's execution time keeps steadily increasing due to the increasing size of its map. Loop closure attempts result in execution time spikes of 100's of ms, an order of magnitude more than its average per-frame execution time.

    {\bf Case Study: VIO}
    \label{subsubsec:results-case-openvins}
        As shown in Table~\ref{table:openvins}, OpenVINS consists of 7 tasks, each with their own compute and memory characteristics.
        The compute patterns span stencils, GEMM, Gauss-Newton, and QR decomposition, among others.
        The memory patterns are equally diverse, spanning dense and sparse, local and global, and row-major and column-major accesses.
        These properties manifest themselves in microarchitectural utilization as well; e.g.,
        the core task of feature matching, KLT, has an IPC of 3, which is significantly higher than the component average of 2.2 due to its computationally intensive integer stencil.
        Studying the tasks of VIO at an algorithmic level suggests that separate accelerators may be required for each task. However, analyzing the compute primitives within each task at a finer granularity shows that tasks may be able to share programmable hardware; e.g., MSCKF update and SLAM update share compute primitives, but have different memory characteristics, thereby requiring a degree of programmability.
        Furthermore, these accelerators must be designed to handle significant variations in execution time.
        
    {\bf Case Study: Audio Playback}
    \label{subsubsec:results-case-playback}
        Audio playback is an example of a component for which building a unified accelerator is straightforward.
        As shown in Table~\ref{table:playback}, we identified 4 tasks in audio playback.
        Psychoacoustic filter and binauralization have \textbf{identical} compute and memory characteristics, while rotation and zoom have similar (but not identical) characteristics.
        Consequently, a traditional audio accelerator with systolic arrays, specialized FFT blocks, and streambuffers should suffice for all four tasks.
        

    {\bf Key implications for architects}
      The diversity of components and tasks, the lack of a single dominant task for most components, and per-frame variability pose several challenges for hardware specialization. It is likely impractical to build a unique accelerator for every task given the large number of tasks and the severe power and area constraints for XR devices: leakage power will be additive across accelerators, and interfaces between these accelerators and other peripheral logic will further add to this power and area (we identified 27 tasks in Figure~\ref{fig:task-breakdown}, and expect more tasks with new components).
      
      This motivates {\em automated techniques for determining what to accelerate.}
          An open question is what is the granularity of acceleration; e.g., one accelerator per task or a more general accelerator that supports {\em common primitives shared among components.}
        %
        Besides the design of the accelerators themselves, architects must also determine {\em accelerator communication interfaces} and techniques to exploit accelerator level parallelism. 
       
        Finally, although not explicitly studied here, algorithms for the various components continue to evolve and there are multiple choices available at any time (\NAME{} already supports such choices). This motivates {\em programmable accelerators} with the attendant challenges of developing a software stack for such a system. \NAME{} provides a testbed to perform all of the above research in the context of the full system.

\section{Research Enabled by \NAME{}}
\label{sec:research}

Architects have embraced specialization but most research focuses on accelerators for single programs. \NAME{} is motivated by research for specializing an entire domain-specific system, specifically edge systems with constrained resources, high computation demands, and goodness metrics based on end-to-end domain-specific quality of output; e.g., AR glasses, robots/drones, autonomous vehicles, etc.
Specific examples of multi-year research agendas enabled by \NAME{} follow below:

(1) Compiler, design space exploration, and HLS tool flows to automatically select common \NAME{} software tasks to accelerate, generate cross-component/hardware-software co-designed programmable and shared accelerators, generate specialized communication mechanisms between multiple such shared accelerators, design generalized accelerator communication/coherence/consistency interfaces, build code generators, and prototype subsystems on an FPGA for end-to-end QoE-driven evaluations. We have already seen significant power/performance/area/QoE opportunities with end-to-end design that are not apparent in isolated component analysis, and affect the architecture of the system. For example, isolated accelerator designs for scene reconstruction have combined the datapaths for the fusion and raycasting operations to improve throughput~\cite{GautierAlthoff2019}. In a full system, raycasting can be shared with both rendering and spatial reprojection, motivating decoupling of its datapath, efficient accelerator communication mechanisms, and shared accelerator management.

(2) End-to-end approximations. Without an end-to-end system-level QoE-driven view, we either lose too much quality by over approximating or forgo extra quality headroom by under approximating. Consider choosing bit precision for specialized hardware for depth processing. Isolated designs have picked a precision to maximize performance/W while remaining within some error bound~\cite{ChoiKim2013} but without considering full system impact. In a full system, the impact of the low precision on the user’s perception depends on how the error flows in the depth $\rightarrow$ scene reconstruction $\rightarrow$ rendering $\rightarrow$ reprojection $\rightarrow$ display chain (each of the intermediate components may have their own approximations and error contributions). Significant research is required to determine how to compose these errors while co-designing accelerators with potentially different precisions and approximations. \NAME{} enables this.

(3) Scheduling and resource management for \NAME{} to maximize QoE and satisfy the resource constraints, and co-designed hardware techniques to partition the accelerator and memory resources  (these techniques will be applicable to many domains where the task dataflow is a DAG).

(4) Common intermediate representations for such a complex workload to enable code generation for a diversity of accelerators and specialized communication interfaces.

(5) New QoE metrics and proxies for runtime adaptation.

(6) Device--edge server--cloud server partitioning of \NAME{} computation, including support for streaming content and multiparty XR experiences in a heterogeneous distributed system.

(7) Security and privacy solutions within resource constraints while maintaining QoE. This is a major concern for XR, but solutions are likely applicable to other domains.

(8) Co-designed on-sensor computing for reducing sensor and I/O power.

The above research requires and is motivated by the QoE-driven, end-to-end system and characterizations in this paper.
\section{Related Work}
\label{sec:related-work}
    
    There is an increasing interest in XR in the architecture community and several works have shown promising specialization results for individual components of the XR pipeline:
    Processing-in-Memory architectures for graphics~\cite{XieSong2017,XieZhang2019}, video accelerators for VR~\cite{LengChen2018,LengChen2019,MazumdarAlaghi2017,MazumdarMoreau2017}, 3D point cloud acceleration~\cite{XuTian2019,FengTian2020}, stereo acceleration~\cite{FengWhatmough2019}, computer vision accelerators~\cite{ZhuSamajdar2018,TriestNikolov2019}, scene reconstruction hardware~\cite{GautierAlthoff2019}, and SLAM chips~\cite{SuleimanZhang2019,MandalJandhyala2019}.
    However, these works have been for single components in isolation.
    Significant work has also been done on remote rendering~\cite{ZhaoZhang2020,MengPaul2020,LiuHe2020,LiuVlachou2020,BoosChu2016,LiGao2018} and 360 video streaming~\cite{BaoZhang2017,Dasari2020,HouDey2018,QianHan2018}. While these works do consider full end-to-end systems, they only focus on the rendering aspect of the system and do not consider the interplay between the various components of the system.
    Our work instead provides a full-system-benchmark, consisting of a set of XR components representing a full XR workflow, along with insights into their characteristics. Our goal is to propel architecture research in the direction of full domain-specific {\em system} design.
    
    There have been other works that have developed benchmark suites for XR.
    Raytrace in SPLASH-2~\cite{SPLASH2}, VRMark~\cite{VRMark}, FCAT~\cite{NVFCAT}, and Unigine~\cite{UNIGINE} are examples of graphics rendering benchmarks, but do not contain the perception pipeline nor adaptive display components.
    The SLAMBench~\cite{nardi2015introducing, bodin2018slambench2,bujanca2019slambench} series of benchmarks aid the characterization and analysis of available SLAM algorithms, but contain neither the visual pipeline nor the audio pipeline.
    Unlike our work, none of these benchmarks suites look at multiple components of the XR pipeline and instead just focus on one aspect.
    
    Chen et al.~\cite{ChenDai2019} characterized AR applications on a commodity smartphone, but neither analyzed individual components nor considered futuristic components.

\section{Conclusion}
\label{sec:conclusion}

This paper develops \NAME{}, the first open source full system XR testbed for driving future QoE-driven, end-to-end architecture research.
\NAME{}-enabled  analysis shows several interesting results for architects -- demanding performance, power, and quality requirements; a large diversity of critical tasks; significant computation variability that challenges scheduling and specialization; and a diversity in bottlenecks throughout the system, from I/O and compute requirements to power and resource contention.
\NAME{} and our analysis have the potential to propel many new directions in architecture research.
Although \NAME{} already incorporates a representative workflow, as the industry moves towards new frontiers, such as the integration of ML and low-latency client-cloud applications, we envision \NAME{} will continue to evolve, providing an increasingly comprehensive resource to solve the most difficult challenges of QoE-driven domain-specific system design.

\section*{Acknowledgements}

\NAME{} came together after many consultations with researchers and practitioners in many domains: audio, graphics, optics, robotics, signal processing, and extended reality systems. We are deeply grateful for all of these discussions and specifically to the following: Wei Cui, Aleksandra Faust, Liang Gao, Matt Horsnell, Amit Jindal, Steve LaValle, Steve Lovegrove, Andrew Maimone, Vegard \O ye, Martin Persson, Archontis Politis, Eric Shaffer, and Paris Smaragdis.

The development of \NAME{} was supported by the Applications Driving Architectures (ADA) Research Center, a JUMP Center co-sponsored by SRC and DARPA, the Center for Future Architectures Research (C-FAR), one of the six centers of STARnet, a Semiconductor Research Corporation program sponsored by MARCO and DARPA, the National Science Foundation under grant CCF 16-19245, and by a Google Faculty Research Award. The development of \NAME{} was also aided by generous hardware and software donations from Arm and NVIDIA.

\bibliographystyle{IEEEtranS}
\bibliography{references}

\end{document}